\documentclass[journal]{IEEEtran}
\usepackage{etoolbox}
\newlength{\BIOskip}
\usepackage{tikz}
\newcommand{\circled}[1]{%
\tikz[baseline=(char.base)]{
\node[shape=circle,draw,inner sep=1.2pt, line width=0.4pt] (char) {\scriptsize\textbf{#1}};}}
\usepackage{hyperref}

\AtEndEnvironment{IEEEbiography}{\vspace{\BIOskip}}
\IEEEoverridecommandlockouts
\usepackage{cite}
\usepackage{amsmath,amssymb,amsfonts}
\usepackage[ruled,vlined,linesnumbered]{algorithm2e}
\SetKw{End}{end}

\usepackage{algpseudocode}
\usepackage{graphicx}
\usepackage{textcomp}
\usepackage{xcolor}
\usepackage{pifont}
\usepackage{graphicx}
\usepackage{booktabs}
\usepackage{float}
\usepackage{booktabs}
\usepackage{multirow}
\usepackage{pifont}
\newcommand{\cmark}{\ding{51}} 
\usepackage[font=footnotesize]{subfig} 
\def\BibTeX{{\rm B\kern-.05em{\sc i\kern-.025em b}\kern-.08em
    T\kern-.1667em\lower.7ex\hbox{E}\kern-.125emX}}
\begin{document}

\title{\textbf{ORACL}: Optimized Reasoning for Autoscaling via Chain of Thought with LLMs for Microservices\\

}

\author{
    Haoyu Bai\textsuperscript{1},
    Muhammed Tawfiqul Islam\textsuperscript{1},
    Minxian Xu\textsuperscript{2},
    Rajkumar Buyya\textsuperscript{1}
    \thanks{\textsuperscript{1}Quantum Cloud Computing and Distributed Systems (qCLOUDS) Lab, School of Computing and Information Systems, The University of Melbourne, Australia.}
    \thanks{\textsuperscript{2}Shenzhen Institute of Advanced Technology, Chinese Academy of Sciences, China.}
    }

\maketitle

\begin{abstract}

Applications are moving away from monolithic designs to microservice and serverless architectures, in which fleets of lightweight, independently deployable components run on public clouds. Autoscaling serves as the de facto control mechanism for balancing resource utilization and Quality of Service (QoS), and its policies are typically refined based on insights from incident root‑cause analysis (RCA), which exposes the capacity bottlenecks and failure patterns that autoscaling must prevent. They together lead to a combinatorial state space: state‑of‑the‑art controllers are either (i) learned models that require substantial per‑deployment training and remain opaque, or (ii) brittle, hand‑tuned rules that fail to generalize. We therefore consider whether Large Language Models (LLMs) can act as universal few-shot resource allocators that adapt across rapidly evolving microservice deployments.
To achieve this, we propose ORACL, \underline{O}ptimized \underline{R}easoning for \underline{A}utoscaling via \underline{C}hain of Thought with \underline{L}LMs for Microservices, a framework that leverages LLMs’ embedded prior knowledge and chain‑of‑thought (CoT) reasoning to diagnose performance regressions and allocate resources. The system collects run-time telemetry (pods, replicas, CPU/memory, latency/SLOs, and fault signals), transforms them into semantically rich natural language state sketches and invokes an LLM that produces an intermediate, interpretable “allocation and root causes finding CoT.” This CoT first identifies likely root causes, then prunes the search space of candidate actions, and finally recommends resource adjustments with guardrails that respect safety and policy constraints. To make this practical, we curate, profile, and annotate over 80GB of traces and metrics drawn from open‑source microservice workloads; we further specialize the model via instruction‑tuned tokens, supervised fine‑tuning, and reinforcement learning to stabilize and regularize the intermediate CoT. Our system demonstrates that LLM‑guided reasoning can efficiently narrow the action space and issue actionable allocations without deployment‑specific retraining, while exposing human‑auditable rationales suitable for operations workflows.
Experimental evaluations show that our proposed approach identifies root causes with 15\% higher accuracy compared to state-of-the-art models, accelerates training by up to 24×, and improves QoS by 6\% in short-term scenarios.
\end{abstract}

\begin{IEEEkeywords}
LLM-based resource management, Microservices, Root Cause Analysis, Orchestration, Quality of Service
\end{IEEEkeywords}

\section{Introduction}
The emergence of complex microservice architectures represents a transformative shift in software system design and management, significantly impacting the technological landscape. These modern applications, composed of lightweight and scalable containerized microservices, greatly enhance scalability, agility, and resilience. Microservices are typically loosely coupled, deployed across diverse physical nodes globally, and managed by various service providers~\cite{alibaba-cloud-overview}. Some functionalities may even reside on privately hosted residential machines due to privacy concerns. Such deployments present diverse configurations, varying considerably in computational capabilities and network latencies~\cite{xu2024practice}. Users often migrate between different cloud platforms or virtual machine types to optimize for energy efficiency, cost savings, or enhanced service performance. Additionally, microservices evolve rapidly, with frequent functional and structural updates driven by developers striving to maintain competitiveness in dynamic software markets.

QoS in microservice applications, often measured by metrics such as average response time \cite{aws-overview}, is significantly affected by both the number of container instances and the resources allocated to each. Flexible deployment strategies and pay-as-you-go pricing models give users substantial control over resource utilization. However, accurately identifying existing and potential bottlenecks, as well as their root causes, is essential for adjusting resources appropriately. In practice, many operators over-provision resources for specific applications to satisfy service-level objectives (SLOs), but this can lead to substantial resource wastage~\cite{196288}. Such inefficiency increases operational costs, by unnecessarily consuming virtual machine capacity, and contributes to environmental impacts. Consequently, efficient RCA and autoscaling techniques are critical to balancing QoS requirements with budget constraints.

Traditional approaches to cloud incident RCA still rely heavily on on-call engineers (OCEs) manually collecting and correlating heterogeneous telemetry, such as logs, metrics, and traces, from multiple applications and services~\cite{zhang2023system}. In large-scale cloud environments, the sheer volume and heterogeneity of this data make it difficult for OCEs to rapidly identify the small subset of signals that are truly relevant to a given incident, thereby delaying effective diagnosis and remediation. Even where supporting automation exists, the overall process remains labor-intensive and error-prone, as engineers must sift through noisy and sometimes incomplete information, interpret complex system behaviors, and iteratively infer likely root causes. As a result, despite substantial advances in data-driven and artificial intelligence–based techniques for operations and monitoring, OCEs still invest significant manual effort in analyzing incident data and determining the underlying causes of failures~\cite{jiang2020mitigate}.

Current autoscaling methodologies predominantly utilize rule-based systems or deep neural networks with extensive parameterization~\cite{xu2025auto}. Rule-based methods are highly dependent on manually defined rules, which may fail to adequately capture complex interactions within microservices due to vast combinatorial complexities. Conversely, deep learning-based approaches require large training datasets and often take extended periods, sometimes weeks, to converge~\cite{yan2020learning}. The associated training latency can degrade initial system performance, and the data collection process itself can be a burden, typically taking tens of seconds to minutes per allocation. Changes to deployment environments or the introduction of new deployments require redesigning neural networks, as shifts in state and action spaces require prolonged retraining. Additionally, these resource-intensive neural network models typically demand significant GPU resources, creating accessibility challenges for small-scale enterprises.

Moreover, existing technologies often overlook significant domain-specific prior knowledge, such as YAML specifications, microservice topologies, service functionalities, relationships within service chains, and specific user requirements. Understanding domain-specific characteristics, for example, higher I/O and memory demands in Online Transaction Processing (OLTP) systems, CPU-intensive operations in Online Analytical Processing (OLAP) applications~\cite{shen2023bridging}. However, current methodologies generally neglect this critical knowledge, thereby relying on inefficient exploratory processes driven purely by neural networks. Leveraging human domain expertise can substantially reduce exploration complexity, leading to more efficient and practical autoscaling and RCA solutions.

To overcome these limitations and leverage recent advancements in LLMs~\cite{openai2025gpt5, qwen3team2025qwen3, anthropic2025claude4}, we propose a semantic-aware resource allocation framework that integrates various cloud-oriented tools, allowing the LLM agents~\cite{he2024emerged} to use different cloud-native tools for automated reasoning and resource optimization. Our method comprehensively models microservice characteristics and employs cloud-specific chain-of-thought reasoning techniques~\cite{wei2022chain} to optimize resource allocation more effectively. 

The main \textbf{contributions} of this paper are:

\begin{enumerate}
\item We design a novel framework that enables LLMs to directly interact with cloud clusters by transforming raw cloud states into semantically meaningful representations, allowing the model to perform automated reasoning and resource adjustment. Based on this design, we construct and annotate a large-scale prompt–response dataset (over 80 GB) that captures engineers’ expertise in diagnosing and analyzing cloud-related issues, serving as a foundation for training and evaluation.

\item We propose a unified algorithm that connects RCA and Resource Allocation through an analogy-based formulation. This approach embeds a training pipeline that optimizes the model’s intermediate reasoning process itself, enabling the framework to generalize across diverse microservice-based applications.

\item We conduct comprehensive testing and validation of the proposed model and policies using real-world traces and a dedicated experimental platform. These evaluations thoroughly assess the effectiveness and performance of both RCA and autoscaling for microservices.
\end{enumerate}

The rest of the paper is organized as follows.  
Section~\ref{sec:relatedworks} provides background and related work on both RCA and autoscaling for microservices.  
Section~\ref{sec:Motivation} discusses the limitations of prior work and provides an illustrative experiment highlighting these shortcomings, thereby motivating a system architecture designed to overcome them.
Section~\ref{sec:SystemDesign} introduces the overall architecture and high-level design of each component. 
Section~\ref{sec:Implementation} details the implementation and underlying mathematical formulations.  
Section~\ref{Experiment} presents a comprehensive experimental evaluation. 
Section~\ref{sec:Limitation} concludes our work and outlines potential directions for future research.
\section{Related Work}
\label{sec:relatedworks}

\begin{table*}[t]
\centering
\caption{A comparison of related works with our proposed system}
\renewcommand{\arraystretch}{1.2}
\resizebox{\textwidth}{!}{
\begin{tabular}{|c|c|c|c|c|c|c|c|c|}
\hline
\multirow{2}{*}{\textbf{Framework}} & 
\multicolumn{2}{c|}{\textbf{Task-Level}} & 
\multicolumn{2}{c|}{\textbf{Data-Level}} & 
\multicolumn{2}{c|}{\textbf{Reasoning-Level}} & 
\multicolumn{2}{c|}{\textbf{System-Level}} \\ \cline{2-9}

 & Root Cause/ Bottleneck & Scheduling & Unstructured Data & Interpretability & Reasoning & Optimized CoT & Unifying Tasks & Transferable \\ 
\hline
Murphy~\cite{harshavardhan2023murphy}                   & \cmark &         &         & \cmark & \cmark &         &         &            \\ \hline
BARO~\cite{pham2024baro}                     & \cmark &         &         & \cmark & \cmark &         &         &            \\ \hline
CoE~\cite{yao2024coe}                      & \cmark &         & \cmark  & \cmark & \cmark &         &         &            \\ \hline
RCACopilot~\cite{chen2024rcacopilot}               & \cmark &         & \cmark  & \cmark & \cmark &         &         & \cmark     \\ \hline
SLIM~\cite{ren2024slim}                     & \cmark &         &         & \cmark & \cmark &         &         &            \\ \hline
DRPC~\cite{bai2024drpc}                            &        & \cmark  &         &        &        &         &         &            \\ \hline
FIRM~\cite{qiu2020firm}                            & \cmark & \cmark  &         &        &        &         & \cmark  &            \\ \hline
AWARE~\cite{qiu2023aware}                           & \cmark & \cmark  &         &        &        &         & \cmark  &  \cmark    \\ \hline
IntentContinuum\cite{akbari2025intentcontinuum} & \cmark &         &         &        &\cmark  &         &         &            \\ \hline
Kubeintellect~\cite{kubeintellect-2025}                   & \cmark &         &         &        &        &         &         &            \\ \hline
Analytical ~\cite{zhang2024analytically}    & \  & \cmark  &  \cmark  &\cmark &   \cmark     &         &    &        \\ \hline
\textbf{ORACL (Our Method)}                      &\cmark  & \cmark   & \cmark & \cmark & \cmark &  \cmark & \cmark  & \cmark     \\ \hline

\hline
\end{tabular}}

\label{tab:relatedWorks}
\end{table*}

In this section, we review root-cause diagnosis and orchestration techniques for microservices. Prior studies \cite{zhang2024vision,harshavardhan2023murphy,qiu2020firm,qiu2023aware} demonstrate that accurate diagnosis plays a critical role in microservice-based systems, as it can either guide orchestration decisions or be reformulated directly as a scheduling problem. Existing methods often rely on anomaly-injection datasets to learn mappings from failure patterns to corrective actions, after which the control procedure is recast as a resource-allocation or scheduling task. This workflow highlights the tight coupling between diagnosis quality and orchestration efficiency in microservice environments.

\subsection{RCA Problems}
\color{blue}

\color{black}
In microservice-based cloud systems, RCA serves as the foundation for reliability engineering by enabling precise fault localization, noise reduction, causality-driven autoscaling, and proactive failure containment. Without accurate RCA, resource management and self-healing actions risk reacting to symptoms rather than actual bottlenecks, leading to instability and inefficient resource usage. 
The seminal work Murphy~\cite{harshavardhan2023murphy} introduced a probabilistic graphical model based on a \textit{Markov Random Field (MRF)} to diagnose performance anomalies from loosely defined metric associations, eliminating the need for an explicit acyclic dependency graph. To enhance reliability and efficiency, BARO~\cite{pham2024baro} integrates Bayesian Online Change Point Detection (BOCPD) with statistical hypothesis testing, unifying anomaly detection and RCA in a single, robust framework.  

Interpretable causal reasoning has also advanced. Chain-of-Event (CoE)~\cite{yao2024coe} learns weighted dependencies among metrics, logs, and traces through an event-level causal graph, achieving over 85\% top-1 accuracy and producing explicit causal chains for human-aligned interpretability. Compared with Murphy’s on-the-fly inference, CoE leverages historical incident data to improve accuracy on recurring failures. Meanwhile, SLIM~\cite{ren2024slim} introduces a lightweight, interpretable rule-based RCA model designed for imbalanced or novel failure data. SLIM generates concise, human-readable \textit{IF--THEN} rules optimized for F1-score and can adapt online, achieving high interpretability and efficiency for continuously evolving systems.

Recently, RCACopilot~\cite{chen2024rcacopilot} introduced LLMs into Microsoft’s incident management workflow, combining structured telemetry with unstructured data such as alerts and tickets to generate natural-language explanations. Deployed at scale, it demonstrates how RCA can be coupled with human-like reasoning and adaptive summarization.

\subsection{Autoscaling}
In microservice architectures, autoscaling dynamically adjusts the number of service replicas based on real-time workload and performance signals, ensuring resource efficiency, service stability, and consistent quality of experience under fluctuating demand.
\subsubsection{Rule Based and Deep Learning Based Approaches}
Autoscaling for microservices spans rule-based heuristics and learning-based controllers. Threshold mechanisms (e.g., AWS Auto Scaling and the Kubernetes HPA) trigger scale-out/in when utilization crosses manually set limits but are workload-agnostic and error-prone\cite{aws2025ec2autoscaling}. To improve robustness, ProScale~\cite{cheng2023proscale} combines short-horizon Simple Moving Average forecasting with a greedy allocator and a decision-tree regressor to predict the required instance count under bursty traffic\cite{cheng2023proscale}. LSRAM further frames service-level-objective (SLO)–aware provisioning as a convex optimization that minimizes resource cost subject to tail-latency constraints, solved via gradient descent and accelerated by a vector-projection/cosine-similarity approximation for distributing partial SLO budgets\cite{hu2025lsram}.

Learning-based methods cast scheduling as sequential decision making. Elarng applies UCB1 multi-armed bandits to select VM counts based on CPU utilization trends\cite{erlang-eurosys-2024}. GRAF builds hierarchical service graphs and uses workload prediction with graph neural networks for dependency-aware allocation, albeit with non-trivial graph maintenance overhead in highly dynamic settings\cite{graf-ton-2024}. Deep RL systems enrich workload representations: FIRM identifies critical chains with SVMs and optimizes resources via an actor–critic agent\cite{qiu2020firm}, while AWARE adds meta-learning and bidirectional RNN encoders to accelerate convergence\cite{qiu2023aware}. However, their synchronous optimization along critical paths can overreact under asynchronous fluctuations. DRPC addresses this by coupling imitation learning with TD3 and decomposing policies into lightweight node-local submodels, reducing inference cost and improving responsiveness at scale\cite{bai2024drpc}.

\subsubsection{LLM Based Approaches}
LLMs are increasingly applied to autonomic system management across intent-driven control, resource scheduling, and multi-agent orchestration. For passive resource allocation, IntentContinuum and KubeIntellect embed LLMs in runtime loops to monitor signals, diagnose SLO violations, and coordinate remediation across heterogeneous Kubernetes clusters \cite{akbari2025intentcontinuum,kubeintellect-2025}. For proactive scheduling, LLMs act as priors or perform action pruning within reinforcement learning (RL) to stabilize task and job-shop allocation \cite{krishnamurthy2025large}. At the edge, the Edge Computing Power Network (ECPN) shows that LLM-guided offloading and allocation reduce energy, accelerate convergence, and enhance privacy under dynamic conditions \cite{sui2024large}. Beyond single controllers, multi-agent schemes improve scalability; for example, the work in \cite{zhang2024vision} proposes a hierarchical design that couples low-level planner–executor agents with higher-level managers, coordinating via protocols and message queues. ServiceOdyssey integrates runtime state tracking with LLM-driven curriculum design, execution planning, and knowledge curation, enabling progressive self-learning and increasingly sophisticated microservice orchestration \cite{yu2025enabling}.

\subsection{Critical Analysis}
\label{sec:criticalAnalysis}
Although representative methods have made valuable contributions, notable limitations remain. Existing methodologies for microservice resource allocation have made significant advancements but exhibit limitations, particularly concerning flexibility, adaptability, and interpretability. Rule-based approaches rely on manually configured thresholds and trigger conditions, inherently limiting their effectiveness due to the combinatorial complexity and NP-hard nature of autoscaling problems. Human-defined rules struggle to encompass the exhaustive permutations and interactions among diverse microservices, leading to suboptimal decisions and restricted applicability across varied scenarios.

Neural networks and reinforcement learning methods similarly face critical constraints, primarily due to their rigid structural designs and heavy reliance on substantial training datasets. These methods necessitate extensive model redesign and retraining whenever microservice configurations evolve, posing significant challenges during frequent updates or early deployment stages. Furthermore, slow convergence rates in actor-critic frameworks restrict rapid adaptation to changes in user behavior or cloud environments. The "black-box" nature of deep learning models additionally diminishes interpretability, reducing user trust and limiting transparency in resource allocation decisions.

Recent work demonstrates the promise of LLMs in cloud management, but most existing approaches still depend on superficial prompt parsing and generic pretrained knowledge. As a result, they often fail to capture the nuanced characteristics of cloud workloads, limiting their accuracy in interpreting complex contextual signals. This reliance on static pretraining data and fixed deduction patterns results in poor generalization to dynamic and heterogeneous cloud environments; the underlying motivations are elaborated in Section~\ref{sec:Motivation}.

\section{Motivation and System Architecture}\label{sec:Motivation}

In this section, we present a motivating experiment, derive the key obstacles it reveals, and outline a high‑level architecture to address them.

\begin{figure}[h]
    \centering
    \includegraphics[width=0.8\linewidth]{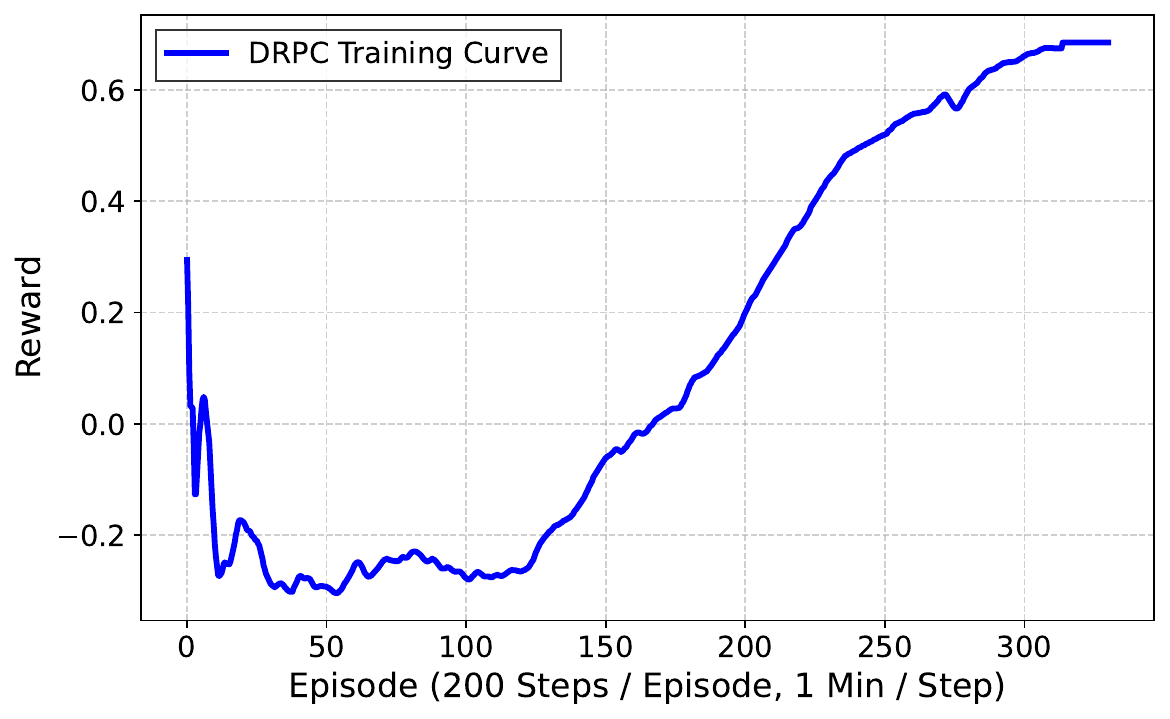}
    \caption{Convergence of the DRPC Algorithm}
    \label{fig:drpc_convergence}
\end{figure}

Fig.~\ref{fig:drpc_convergence} reports the convergence behavior of a representative DRPC trained on the Train‑Ticket microservice application~\cite{gan2019open}. The application is deployed on a five‑node cluster (one master and four workers); each node has 8~vCPUs and 8~GB of memory. The deployment comprises approximately 40 microservices (e.g., \textit{user}, \textit{station}, \textit{price}, \textit{route}). The reward jointly considers resource utilization and latency. Without synthetic‑data augmentation, roughly 19~days of traces are required for DRPC to reach usable performance.

Guided by this observation and the analysis in Sec.~\ref{sec:criticalAnalysis}, we target two core obstacles: (i) limited generality under topology churn and (ii) poor sample efficiency at microservice scale. We additionally aim to (iii) improve interpretability and (iv) unify RCA and orchestration.

Our approach draws inspiration from two sources: (i) DeepSeek‑R1~\cite{deepseek2025r1}, which explicitly optimizes intermediate reasoning steps, and (ii) the question‑driven, decision‑tree‑like workflow that cloud engineers typically follow when diagnosing complex environments.

The key idea is to encode application‑agnostic microservice semantics in natural language that mirrors how practitioners diagnose and act. Because a microservice call‑graph can be expressed textually, the representation is \emph{universal}, rather than tied to a single, static application. By designing a structured output schema, we unify RCA and orchestration within one loop and treat the reasoning procedure itself as the decision policy—analogous to optimizing the rule set in a rule‑based system. This design yields interpretable rationales, greater robustness to configuration and topology changes, and reduced sample complexity and training cost.

\begin{figure}[h]
    \centering
    \includegraphics[width=0.85\linewidth]{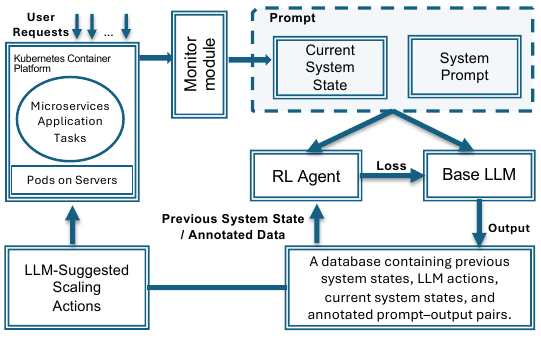}
    \caption{ORACL System Architecture}
    \label{fig:ArchitectureRaj}
\end{figure}

Building on these ideas, we propose \textbf{ORACL} and its high-level architecture as shown in  Fig.~\ref{fig:ArchitectureRaj}.  Cluster telemetry (e.g., latency, requests per second) is normalized and appended to a system prompt that concisely describes the deployment and relevant call-path context. The aggregated prompt is then provided to an LLM, which returns a recommended resource allocation or remediation action; then, the action is executed on the cluster. In parallel, a reinforcement learning (RL) agent collects annotated prompt--output pairs together with historical and current system states. Each executed action induces a new system state, which we merge with the prior state to form a state transition. These transitions, along with the annotated prompts, constitute the training data used to fine-tune the base LLM, thereby closing the loop among diagnosis, decision making, and learning.

\begin{figure*}[!t]
  \centering
  \includegraphics[width=0.7\textwidth]{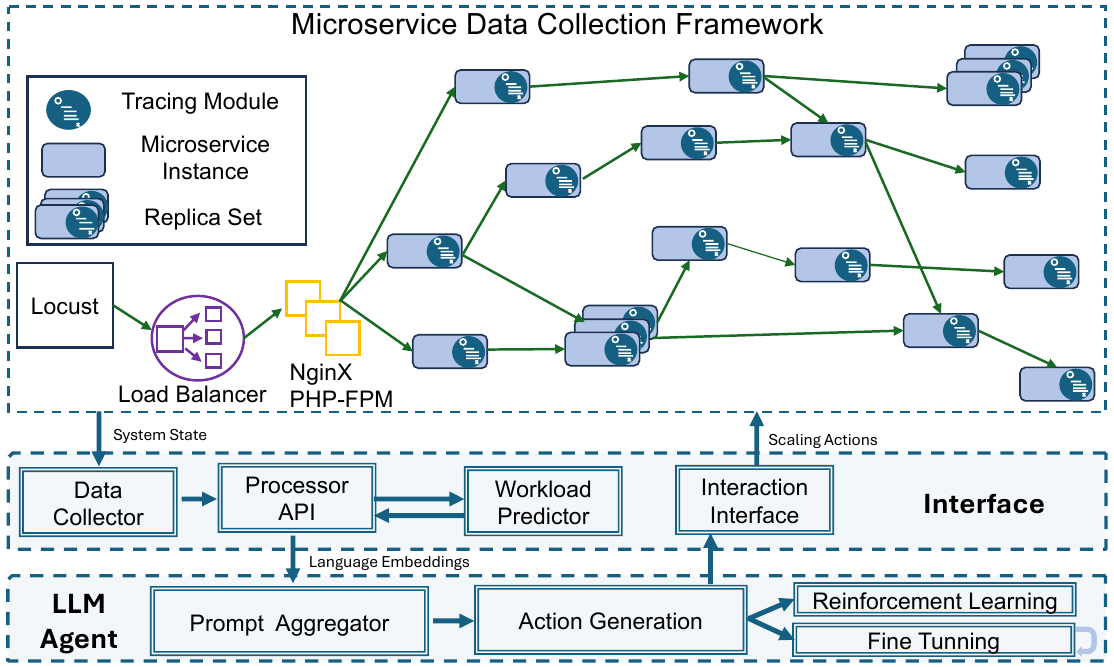}
  \caption{ORACL System Prototype Diagram}
  \label{fig:SystemArchitecture}
\end{figure*}
\section{System Design}\label{sec:SystemDesign}
In this section, we present a system design for our approach, organized according to the Monitor–Analyze–Plan–Execute framework over a shared knowledge base (MAPE‑K), as illustrated in Fig.~\ref{fig:SystemArchitecture}. The system comprises four primary components:
(1) a tracing module that continuously observes and records system state;
(2) an interface module that exposes essential capabilities the LLM cannot directly perform;
(3) an LLM agent that aggregates contextual evidence to produce structured outputs and incorporates a reinforcement‑learning module to support both online and offline learning.

\subsection{Interface Module}\label{sec:interface}
Given the practical limitations of LLMs, the interface module orchestrates a three-stage workflow. \textbf{Data Collector} captures and normalizes system telemetry and call-path context. \textbf{Workload Predictor} forecasts near-term workload based on these observations. \textbf{Processor API} fuses the signals into a structured \emph{system-state prompt} and delivers it to the LLM agent. The \textbf{Interaction Interface} then verifies the returned actions before execution.

\subsubsection{Data Collector}\label{subsec:data-collector}
LLMs have limited numerical sensitivity; for example, they may struggle to reliably distinguish small but important differences in metrics (e.g., CPU utilization of 80\% vs.\ 85\%). We therefore discretize continuous values into qualitative bins (e.g., \emph{Low}/\emph{Medium}/\emph{High}) and encode derived efficiency indicators using the same scheme. LLMs also lack an intrinsic notion of healthy versus abnormal states, so we attach explicit thresholds or statistical baselines to guide reasoning. Finally, because LLMs cannot directly observe real‑time behavior, We integrate external monitoring tools (e.g., Prometheus for metrics and Jaeger for tracing\footnote{Prometheus: \url{https://prometheus.io}; Jaeger: \url{https://www.jaegertracing.io}}) to provide structured inputs for LLM-based decision making.

\subsubsection{Workload Prediction Unit}\label{subsec:workload-pred}
Real‑world request streams exhibit strong temporal and spatial correlations, making workloads often predictable~\cite{luo2022power}. However, accurate time‑series prediction remains challenging for general‑purpose LLMs without substantial domain‑specific fine‑tuning. To avoid entangling numeric forecasting with higher‑level reasoning, we offload forecasting to an external API~\cite{luo2022power} dedicated to workload prediction and supply only the prediction (and, when available, its confidence) to the prompt.

\subsubsection{Interaction Interface}\label{subsec:Interaction-Interface}
Because the LLM cannot directly manipulate cluster resources, all actions are mediated by an external execution API. Despite fine-tuning and reinforcement-learning–based optimizations, hallucinations and other high-confidence yet incorrect or conflicting outputs may still occur. To mitigate risk, we apply a simple rule‑based verification layer before execution that rejects obviously unsafe decisions and thus helps preserve system reliability.

\subsection{LLM Agent}\label{sec:llm-agent}
This clause specifies the LLM Agent. The Agent comprises: (i) a Prompt Aggregation Module (PAM) that encodes objectives, constraints, and context into a structured prompt; (ii) an Action‑Generation Module (AGM) that converts LLM outputs into executable actions for the interface; and (iii) a Reinforcement‑Learning and Fine‑Tuning module (RLFT) that implements a two‑stage training procedure.

\subsubsection{Prompt Aggregator}
The PAM consolidate system objectives, temporal constraints, and user policies into prompts suitable for LLM reasoning. The prompts should (a) encode guidance policy and QoS/SLO constraints, (b) carry explicit microservice context, and (c) specify a machine‑readable output schema to enable auditable and reproducible decisions.

\begin{figure}[t]
    \centering
    \includegraphics[width=1\linewidth]{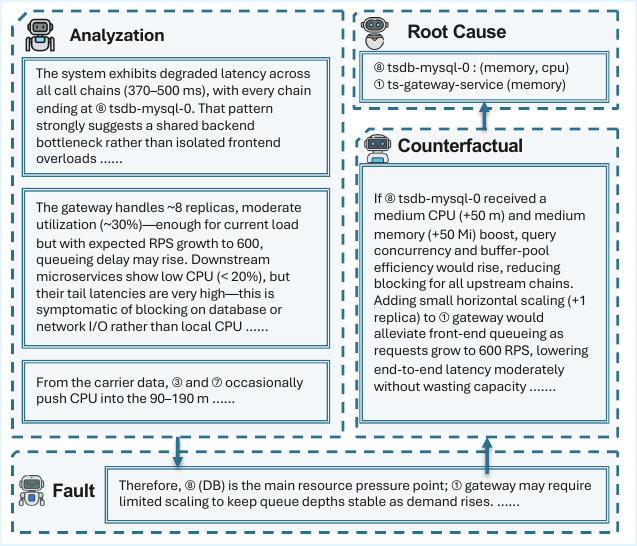}
    \caption{An Expected Output From LLM}
    \label{fig:ExpectedOutput}
\end{figure}

\begin{enumerate}

\item Initial Reasoning Guidance: The PAM provides a reasoning template that directs the LLM to assess interdependent system states systematically before proposing scheduling actions. The template requires explicit intermediate considerations and specify which aspects to cover during CoT reasoning. This structure may improve decision accuracy, stability, and interpretability by making each reasoning step traceable, especially in edge cases.

\item Deployment Description: For each deployment, the PAM accept a natural-language descriptor specifying its functional role, resource-demand profile, runtime behavior, and \emph{readiness time} (time to become operational). Readiness time guides orchestration decisions, services with long readiness times \emph{should} be scale-out earlier to preserve responsiveness and efficiency.

\begin{table}[t]
    \centering
    \renewcommand{\arraystretch}{1.1}
    \setlength{\tabcolsep}{3pt}
    \resizebox{0.8\linewidth}{!}{%
    \begin{tabular}{|c|p{0.6\linewidth}|}
    \hline
    \textbf{Symbol} & \textbf{Explanation} \\
    \hline
    {\centering
    \circled{C}\ \circled{D}\ \circled{M}\ \circled{N}\par}

    & CPU / Disk / Memory / Network-intensive, downstream (→) \\
    \hline
    {\centering
    \circled{c}\ \circled{d}\ \circled{m}\ \circled{n}\par} 
    & CPU / Disk / Memory / Network-intensive, upstream (←) \\
    \hline
    \textcircled{1} -- \textcircled{20} & Microservice identifiers\\
    \hline
    \texttt{<think>} & generate hypotheses \\
    \hline
    \texttt{<Fault>} & list possible causes \\
    \hline
    \texttt{<Counterfactual>} & effect if resources change \\
    \hline
    \texttt{<root>} & root cause: IDs and type \\
    \hline
    \texttt{<Carrier>} & effectiveness of resources change \\
    \hline
    \end{tabular}%
    }
    \caption{Resource and CoT symbol definitions}
    \label{tab:symbols}
\end{table}

\item Call Graph Structure: The call graph is central to reasoning about dependency structure and fan-in/fan-out. It exposes critical paths, bottlenecks, latency‑sensitive paths, and failure propagation to inform resilience and resource allocation. We model static topology and dynamic traces in natural language with attention mechanism and use compact special tokens (Table~\ref{tab:symbols}) to preserve structure and reduce space.

\item Expected Generation Results: The PAM defines a structured output schema; all annotations and pipeline outputs conform to it (Fig.~\ref{fig:ExpectedOutput}).  We treat RCA and resource allocation as distinct yet tightly coupled tasks that share evidence (metrics, traces, and fault) and reasoning. We use a four‑stage pipeline per service per analysis window: (1) Analyzation: summarize salient signals; (2) Fault Identification: list anomalies and candidate symptoms; (3) Counterfactual Reasoning: assess whether resource changes would mitigate degradation; (4) Root‑Cause Resolution: record validated causes and remediations. Symbols are given in Table~\ref{tab:symbols}.

\item Cluster Running State: As outlined in Section~\ref{sec:interface}, this field is paired with a \texttt{Carrier} that encodes system‑level sensitivity and specifies how the discretized state responds to load and configuration changes.
\end{enumerate}

\subsection{Action Generation}
The action-generation module converts LLM outputs into actions that the interface can execute. For RCA, it parses root causes enclosed in \texttt{<root>} tags. For resource allocation, it uses the counterfactual reasoning segment to extract configuration changes, selects those predicted to improve outcomes, and emits an executable action specification.

\subsection{Reinforcement Learning and Fine-Tuning Module}\label{sec:rlft}
Our framework optimizes the reasoning process rather than low-level control actions, enabling the model to learn high-level semantics and generalizable allocation rules. Supervised fine-tuning~\cite{dong2024abilities} trains on annotated prompt-output pairs to enforce the structured schema, encode practical engineers' heuristics, and reasoning patterns. The reinforcement learning component operates in two distinct stages: (i) root-cause identification, optimized with causal analysis and validity checks; and (ii) prediction of whether proposed counterfactual changes achieve the intended system behaviour.

\label{Generation}

\section{System Modeling}\label{sec:Implementation}
This section defines the problem formulation and optimization objective, describe system-wide operation, and present the algorithmic design used to achieve the objective.

\subsection{Problem Formulation}
In ORACL, we formalize resource management for microservices running on a cluster of physical machines. Let $M = \{ m_1, m_2, \ldots, m_i \}$ denote the set of machines. For each machine $m_i \in M$, the allocatable resources are represented as $R_{m,j} = \{ r_{m,1}, r_{m,2}, \ldots, r_{m,j} \}$, where each $r_{m,j}$ denotes a resource type $j$ such as CPU or memory. The set of feasible actions is $A_{m,i} = \{ a_{m,1}, a_{m,2}, \ldots, a_{m,i} \}$, where each $a_{m,i}$ encodes a scaling operation $i$ (horizontal scaling, CPU vertical scaling, or memory vertical scaling) on machine $m$. We denote the textual input provided to the system as $Prompt$, and the generated response sequence of the LLM as $(g_1, g_2, \ldots, g_T)$.

Our methodology consists of two interconnected sub-problems. The first is \textbf{root cause identification}, which we treat as a classification problem. Given input signals $Prompt$, the model predicts $\hat{y} = f_{\theta}(Prompt)$, aiming to maximize the probability $\mathbb{P}(\hat{y} = y^\ast \mid Prompt)$, where $y^\ast$ is the true underlying cause. 

The second sub-problem is \textbf{resource allocation}, which we formulate as a Markov Decision Process (MDP). At each step $t$, the system observes a state $s_t \in \mathcal{S}$, takes an action $a_t \in \mathcal{A}$, and transitions to a new state $s_{t+1}$. The objective is to maximizes the QoS, which can be naturally achieved by letting the LLM perform proactive counterfactual reasoning: given a hypothetical intervention $\tilde{a}_t$, which normally corresponds to the action suggested after identifying a potential problem, the model predicts a future state $\hat{s}_{t+1} = f_{\theta}(s_t, \tilde{a}_t)$ and seeks to maximize the alignment $\mathbb{P}(\hat{s}_{t+1} \approx s_{t+1} \mid s_t, \tilde{a}_t)$. This design ensures that “what-if” predictions guide scaling actions while maintaining consistency with observed system performance.

In summary, ORACL decomposes the learning problem into two stages that share the same chain-of-thought policy~$\pi$, whose conceptual form is expressed in~\eqref{eq:oracl_obj}:
\begin{equation}
\pi^\ast \propto
\underbrace{\mathbb{P}(\hat{s}_{t+1} = s_{t+1} \mid Prompt)}_{\text{counterfactual consistency}}
\underbrace{\mathbb{P}(\hat{y} = y^\ast \mid \hat{s}_{t+1},\, Prompt)}_{\text{root cause classification}}.
\label{eq:oracl_obj}
\end{equation}

\subsection{System-wide Actions}

\begin{algorithm}[t]
\footnotesize

\caption{System-wide Operation Procedure}
\label{algo:System-wide_operation}
\KwIn{
Flags: $Required\_offline\_Training$, $Required\_Online\_Training$;\\
Data instance: $data\_instance$; Task type: $Task\_type$;\\
Base model: $base\_LLM$; 
}
\KwOut{System-level output (e.g., root cause or scheduling actions)}

\BlankLine
\textbf{Initialization:}\\
Load $base\_LLM$\;
\If{$Required\_offline\_Training$}{
    \texttt{Offline\_Training($data\_instance$, $base\_LLM$)}\;
    \End\;
}

\BlankLine
\textbf{// Root-Cause Identification}\\
\If{$Task\_type = ``Root\_cause\_Identification"$}{
    $system\_state \leftarrow get\_cluster\_state()$\;
    \If{$system\_state$ is \textbf{None}}{
        $system\_state \leftarrow data\_instance$\;
        \End\;
    }
    $prompt \leftarrow system\_state$\;
    $Root\_cause$ $\leftarrow base\_LLM.generate(prompt).extract\_between\_tags()$\;
    
    \KwRet{$Root\_cause$}\;
    \End\;
}

\BlankLine
\textbf{// Dynamic Allocation Loop}\\
\If{$Task\_type = ``Allocation"$}{
    \While{True}{
        \texttt{Online\_Training($base\_LLM$, $Required\_Online\_Training$)}\;
        sleep($interval\_time$)\;
    }
    \End\;
}
\end{algorithm}

\textbf{Operation procedure:} In ORACL, the dual-objective optimization problem is formulated as a unified system that operates in two distinct modes. As shown in Algorithm~\ref{algo:System-wide_operation}, we first initialize the base LLM (lines~1--2). When offline training is required, the procedure follows Algorithm~\ref{algo:Offline} (lines~3--5). After the offline training phase, if the task is labeled as an RCA problem, the current system state—either retrieved from historical records or collected from an online cluster—is transformed by the prompt aggregator described in the previous section. The aggregated prompt is then passed to the LLM (lines~6--12). The LLM produces outputs structured into four segments corresponding to the CoT reasoning process, from which the root causes are extracted by parsing the contents enclosed in the \texttt{<root>} tags (lines~13--15). In contrast, when the task is specified as scheduling (lines~16--18), the LLM manager enters a continuous loop in which it proactively allocates resources according to Algorithm~\ref{algo:online}, followed by a sleep interval (lines~20--21). In practical deployments, multiple LLM managers may run in parallel to provision resources for pods associated with the same call graph, thereby improving system-level efficiency. The two algorithms referenced above are elaborated in subsequent sections.

\begin{figure}[H]
    \centering
    \includegraphics[width=1\linewidth]{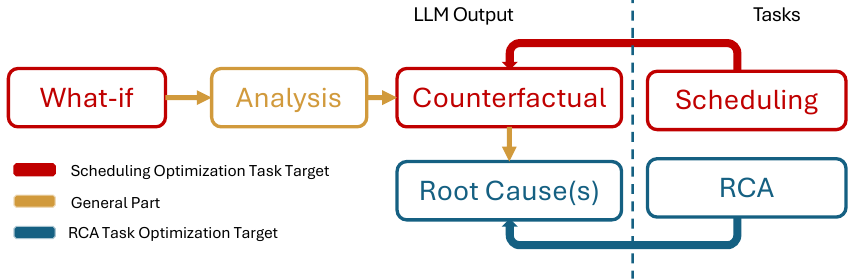}
    \caption{System-wide operation with a Dual Objective Procedure}
    \label{fig:Procedure}
\end{figure}

\textbf{Optimization targets:} As illustrated in Fig.~\ref{fig:Procedure}, the RCA objective is to accurately identify all faulty components (blue), with precision as the primary evaluation metric. For resource allocation, only counterfactual configurations (red) with positively predicted outcomes are retained, ensuring consistency with the CoT-based reasoning process. By aligning counterfactual selection with accurate system-response predictions (yellow), the resulting decisions naturally drive improvements in overall system performance. Discrepancies between predicted and observed outcomes are further fed back to refine the reasoning process and enhance robustness over time~\cite{guo2025deepseek}.

\subsection{Root-Cause Identification}
\begin{algorithm}[t]
\small
\caption{ORACL: Offline Training}
\label{algo:Offline}
\KwIn{
  Annotated dataset: $\mathcal{D}$; Base model: $base\_LLM$;\\
Number of supervised steps: $SFT\_steps$; Number of GRPO steps: $GRPO\_steps$
}
\KwOut{Updated model parameters $\theta$}

\BlankLine
\textbf{Initialization:}\\
Split $\mathcal{D}$ into $(data\_partition_0, data\_partition_1)$ using \texttt{Annotated\_Data.split()}\;
Initialize optimizer and learning scheduler for $base\_LLM$\;

\BlankLine
\textbf{// Supervised Fine-Tuning (SFT)}\\
\For{$step \leftarrow 1$ \KwTo $SFT\_steps$}{
    $logits \leftarrow base\_LLM(data\_partition_0.inputs)$\;
    $loss \leftarrow CrossEntropy(logits,\ data\_partition_0.response)$\;
    $update(base\_LLM,\ loss)$\;
}

\BlankLine
\textbf{// GRPO Fine-Tuning}\\
\For{$step \leftarrow 1$ \KwTo $GRPO\_steps$}{
    $out \leftarrow base\_LLM.generate(data\_partition_1.prompts)$\;
    $[Observation,\ Fault,\ Counterfactual,\ Root\_cause] \leftarrow out.extract\_between\_tags()$\;
    $loss \leftarrow Overall\_Loss(Root\_cause,\ data\_partition_1.root\_cause)$\;
    $update(base\_LLM,\ loss)$\;
}
\end{algorithm}

Algorithm~\ref{algo:Offline} illustrates the offline training procedure for RCA. The annotated dataset, formatted as described in the previous section, is first divided into two partitions (lines~1--2), and the optimizer together with the base LLM parameters is initialized (line~3). The first partition is used to enable the model to not only retain its pretrained knowledge but also adapt to the reasoning patterns commonly employed by cloud engineers when diagnosing faults and proposing solutions.

\begin{equation}
\mathcal{L}(\theta) = - \sum_{t=1}^{T} \log P_{\theta}\!\left(y_t \mid y_{<t}\right)
\label{eq:ntp_loss}
\end{equation}

Subsequently, the entropy loss is calculated as defined in~\eqref{eq:ntp_loss}, and the model parameters are updated accordingly (lines~5--8). The model then proceeds to the critical stage, where reinforcement learning is applied to optimize the CoT process itself. In this stage, the LLM generates four reasoning steps, and the outputs are evaluated by a reward function in conjunction with the annotated data. The design and objective of this reward function are discussed in detail in Section~\ref{sec:offlineObjective}.

\subsubsection{Objective}
\label{sec:offlineObjective}
The \textbf{primary objective} is to maximize the probability of correctly identifying the root cause in~\eqref{eq:rootcauseformula}:

\begin{equation}
\pi^{*} 
= \arg\max_{\pi}\;
\mathbb{E}\Bigg[
P_{\pi}\!\left(
t_{n+1},t_{n+2},\ldots,t_{m}\;\middle|\;
t_{1},t_{2},\ldots,t_{n}
\right)
\Bigg].
\label{eq:rootcauseformula}
\end{equation}

{\small
\noindent
where
\[
t_{1},\ldots,t_{n} \in \mathcal{T}_{\text{reason}},
\quad
t_{n+1},\ldots,t_{m} \in \mathcal{T}_{\text{rc}},
\]
denote the reasoning token sequence and the root cause token sequence, respectively.
}

The \textbf{secondary objective} is to maximize the probability of uncovering false positives during reasoning as shown in~\eqref{eq:secondaryobjective}:

\begin{equation}
\max_{\pi}\;
\mathbb{P}_{\pi}\!\Big(
t^{(\mathrm{cf})}_{1},\ldots,t^{(\mathrm{cf})}_{n}
\;\Big|\;
t^{(\mathrm{cand})}_{1},\ldots,t^{(\mathrm{cand})}_{m}
\Big),
\label{eq:secondaryobjective}
\end{equation}

{
\noindent
where $t^{(\mathrm{cand})}_{1:m}\in\mathcal{T}_{\mathrm{cand}}$ 
denote the candidate root-cause tokens already generated, and 
$t^{(\mathrm{cf})}_{1:n}\in\mathcal{T}_{\mathrm{cf}}$ 
denote the subsequent counterfactual reasoning tokens that may revise or reject earlier candidates.
}

This step ensures that the model not only identifies faulty components with high accuracy but also validates its own reasoning by aligning counterfactual predictions with observed system behavior. In practice, this coupling reduces hallucinations in LLM outputs, improves the robustness of root cause identification, and enhances the reliability of subsequent resource allocation decisions, thereby aligning learning objectives with the overarching goals of SLO compliance and efficient cloud resource use.

\subsubsection{Reward Functions}
The reward functions address correctness, structural properties, length, and over-shifting behavior of the model, as described below.

\paragraph{Correctness}
The correctness score comprises three components: pod-level accuracy, root-cause correctness, and a thresholded penalty that discourages excessive guessing.

\textbf{Pod-level match (guess reward).}
The pod-level match score \(S_{\text{pod}}\) is defined as
\begin{equation}
\label{eq:Spod}
S_{\text{pod}}
= \frac{\mathrm{POD}_{\text{match}}}
       {\mathrm{POD}_{\text{match}} + \mathrm{POD}_{\text{not}} + \epsilon},
\end{equation}
where \(\mathrm{POD}_{\text{match}}\) and \(\mathrm{POD}_{\text{not}}\) denote the numbers of correctly and incorrectly predicted pods, respectively, and \(\epsilon>0\) is a small constant for numerical stability. As shown in~\eqref{eq:Spod}, this term rewards correctly guessed faulty pods irrespective of the specific CPU/Memory/Disk dimension.

\textbf{Root-cause correctness.}
Given the numbers of true positives, false positives, and false negatives, denoted by \(\mathrm{TP}\), \(\mathrm{FP}\), and \(\mathrm{FN}\), respectively, the root-cause correctness is measured by the \(F_\beta\) score:
\begin{equation}
\label{eq:Fbeta}
F_\beta
= \frac{(1+\beta^2)\,\mathrm{TP}}
       {(1+\beta^2)\,\mathrm{TP} + \beta^2\,\mathrm{FN} + \mathrm{FP} + \epsilon},
\qquad \beta>1 ,
\end{equation}
where \(\beta>1\) biases the metric toward recall to reduce missed root causes. As in ~\eqref{eq:Fbeta}, false negatives are weighted more heavily than false positives.

\textbf{Base reward.}
The base reward \(R_{\text{base}}\) combines the pod-level match score in~\eqref{eq:Spod} and the root-cause correctness in \eqref{eq:Fbeta}:
\begin{equation}
\label{eq:Rbase}
\begin{aligned}
R_{\text{base}}
&= \alpha\, S_{\text{pod}}
  + (1-\alpha)\, F_\beta \\
&\quad + \delta \cdot \mathbb{I}[\mathrm{TP}>0],
\qquad 0<\alpha\ll 1,\ \delta>0 .
\end{aligned}
\end{equation}

where \(\alpha\) is a small weight on the pod-level term and \(\delta\) provides an additional bonus whenever at least one true positive is identified. Equation~\eqref{eq:Rbase} thus encourages the model to correctly localize the faulty pod while emphasizing root-cause correctness.

\textbf{Thresholded penalty.}
To discourage missing too many true root causes, a thresholded penalty is applied to the base reward in~\eqref{eq:Rbase}. Let \(\tau_{\mathrm{fn}} \in \mathbb{N}\) be the tolerance threshold for false negatives. The final correctness reward \(R_{\text{Result}}\) is given by
\begin{equation}
\label{eq:Rresult}
R_{\text{Result}}
= R_{\text{base}} - d \cdot \max(0, \mathrm{FN} - \tau_{\mathrm{fn}}),
\end{equation}
where \(d > 0\) controls the penalty strength. When \(\mathrm{FN} \le \tau_{\mathrm{fn}}\), \eqref{eq:Rresult} reduces to the base reward in~\eqref{eq:Rbase}. In contrast, false positives \(\mathrm{FP}\) are only softly penalized through the \(F_\beta\) term in~\eqref{eq:Fbeta}, and no explicit threshold is imposed on them.

\paragraph{Format Compliance}
The objective is to ensure that the LLM outputs the required tags correctly.  
If all tag pairs are valid, no penalty is applied.  
Otherwise, each violation reduces the representative score.  
Formally, let $C_{\text{total}}$ denote the total number of checks  
and $C_{\text{invalid}}$ the number of checks not passed.  
We define:
\begin{equation}
R_{\text{format}} = - \frac{C_{\text{invalid}}}{C_{\text{total}}},
\end{equation}
where each of the following counts as a failed check:
\begin{enumerate}
    \item Empty reasoning content between a start and end tag,
    \item Incorrect ordering of start and end tags,
    \item Nested tags inside another tag’s scope.
\end{enumerate}
Thus, the final score decreases proportionally to the number of violations.

\paragraph{Length Constraint with Smooth Decay}
To balance correctness and inference efficiency, we constrain the model's reasoning length within a statistically determined range $[L_{\min}, L_{\max}]$. Although these bounds are derived from empirical measurements, the model typically requires more explicit reasoning steps and exploratory attempts than an expert human engineer. Consequently, $L_{\max}$ is intentionally set to exceed the typical annotation length produced by human experts. As shown in~\eqref{eq:length}, answers shorter than $L_{\min}$ receive the maximum reward; those with lengths between $L_{\min}$ and $L_{\max}$ are linearly down-weighted; and any answer exceeding $L_{\max}$ obtains zero reward.

\begin{equation}
\label{eq:length}
s(L) =
\begin{cases}
0, & L \le L_{\min}, \\[1.2ex]
-\dfrac{L - L_{\min}}{L_{\max} - L_{\min}}, & L_{\min} < L < L_{\max}, \\[2.0ex]
-1, & L \ge L_{\max}.
\end{cases}
\end{equation}

Formally, let $\text{correct}(c)\in\{0,1\}$ be the correctness indicator, the final reward is dipicted in the~\eqref{eq:lengthFinal}:

\begin{equation}
\label{eq:lengthFinal}
R_{\text{length}}(c) = \text{correct}(c) \cdot s(L(c)).
\end{equation}

\paragraph{KL Regularization}
To prevent the new policy from diverging excessively from the reference model and thereby losing the embedded expertise of cloud engineers, a KL penalty term is introduced, as defined in~\eqref{eq:klreward}.
\begin{equation}
\label{eq:klreward}
R_{\text{KL}} = - \beta \, D_{\mathrm{KL}}\!\bigl(\pi_\theta \,\|\, \pi_{\mathrm{old}}\bigr),
\end{equation}
where $\pi_\theta$ denotes the new policy, $\pi_{\mathrm{old}}$ represents the reference policy from the previous step, and $\beta>0$ controls the strength of the regularization.

\paragraph{Final Reward}
The overall reward $R_{\text{total}}$ is defined as the sum of all the aforementioned components, as shown in~\eqref{eq:totalReward}:
\begin{equation}
\label{eq:totalReward}
R_{\text{total}} = R_{\text{result}} + R_{\text{format}} + R_{\text{length}} + R_{\text{KL}}.
\end{equation}

\subsubsection{With GSPO}
As the main issue is the retrieval of datapoints, we allow our LLM to generate multiple instances each round by deploying Group Sequence Policy Optimization (GSPO) \cite{zheng2025group} with the objective function shown in~\eqref{eq:gspoobj}:

\begin{equation}
\begin{split}
J_{\text{GSPO}}(\theta) 
= \mathbb{E}\Bigg[
   \frac{1}{G}\sum_{i=1}^G 
   \min \Big(
      s_i(\theta) \hat{A_i},\;
      \text{clip}\big(s_i(\theta)\big) \hat{A_i}
   \Big)
\Bigg],
\end{split}
\label{eq:gspoobj}
\end{equation}

Here, each prompt $x$ is sampled from the dataset $\mathcal{D}$, and $\{y_i\}_{i=1}^G$ denotes the $G$ candidate outputs generated by the model. The advantage $\hat{A}_i$ for the $i$-th output is computed by normalizing its total reward $R_{\text{total}}(x,y_i)$ with the group mean and standard deviation.

\begin{equation}
s_i(\theta) =
\left( \frac{\pi_\theta(y_i \mid x)}
            {\pi_{\theta_{\text{old}}}(y_i \mid x)} \right)^{\tfrac{1}{|y_i|}}
\label{eq:ratio}
\end{equation}

where the ratio compares sequence-level likelihoods~\cite{zheng-etal-2023-click} under the new policy $\pi_\theta$ and the old policy $\pi_{\theta_{\text{old}}}$, and the geometric mean (the $|y_i|$-th root) helps prevent instability on long sequences, as shown in \eqref{eq:ratio}. Finally, the clipping function $\mathrm{clip}(\cdot)$ bounds $s_i(\theta)$ within $[1-\epsilon, 1+\epsilon]$ to avoid overly large policy updates.

\subsection{Resource Allocation}
\label{sec:resourceallocation}
\begin{algorithm}[h]
\footnotesize
\caption{ORCAL: Scheduling with Optional Online Training}
\label{algo:online}
\KwIn{
    Base model $base\_LLM$; \\
    Online training flag $Required\_Online\_Training$
}
\KwOut{
    Updated model $base\_LLM$ (if online training enabled)
}

\BlankLine
\textbf{Initialization:} \\
$carrier \leftarrow state\_gathering\_phase(previous\_system\_state)$ \;
$previous\_system\_state \leftarrow get\_cluster\_state()$ \;

\BlankLine
\textbf{Prompt Construction and Inference:} \\
$prompt \leftarrow compose\_prompt(previous\_system\_state,\ carrier)$ \;
$output \leftarrow base\_LLM.generate(prompt)$ \;
$[Observation,\ Fault,\ Counterfactual,\ Root\_cause] 
    \leftarrow output.extract\_between\_tags()$ \;

\BlankLine
\textbf{Counterfactual Action Execution:} \\
$apply\_action(Counterfactual)$ \;

\BlankLine
\textbf{Online Adaptation (Optional):} \\
\If{$Required\_Online\_Training$}{
    $new\_system\_state \leftarrow get\_cluster\_state()$ \;
    $Loss \leftarrow is\_Expected(Counterfactual,\ new\_system\_state)$ \;
    $base\_LLM.step(Loss)$ \;
}
\end{algorithm}

The orchestration procedure consists of an inference phase and an optional online training phase. 
As shown in Algorithm~\ref{algo:online}, because the base LLM does not have an explicit notion of a “good’’ system state, 
we prepend a brief statistical state-gathering phase ~\cite{zhang2024analytically} that characterizes how sensitive the system is to configuration changes. Using a lightweight clustering technique~\cite{blundell2001estimation}, historical system states are summarized into a compact representation, referred to as the \emph{carrier} (lines~1–2). The carrier is then prepended to the current cluster state, and the combined information is encoded into a prompt (lines~3–5).  

The LLM prediction is parsed by an interpreter, defined in~\eqref{eq:interpreter}, to extract the counterfactual action that is expected to lead to a positive change in the system:
\begin{equation}
\label{eq:interpreter}
\hat{o}, a = \pi_{\text{Interpreter}}(\text{Counterfactual}),
\end{equation}
where $\pi_{\text{Interpreter}}(\cdot)$ can be instantiated as either another LLM or a rule-based procedure. The resulting action $a$ is then applied to the managed application (lines~6–9).

If online training is enabled, the true outcome $o$ is observed from the cluster after executing an scaling action $a$ (lines~10–12). We then compute a simple binary loss, analogous to that used in DeepSeek-Math (lines~13), as given in~\eqref{eq:deepseek_loss}. Specifically, the loss is set to 1 if the predicted outcome $\hat{o}$ matches the observed outcome $o$, and 0 otherwise:
\begin{equation}
\label{eq:deepseek_loss}
\mathcal{L}(\hat{o}, o) =
\begin{cases}
1, & \text{if } \hat{o} = o, \\
0, & \text{otherwise}.
\end{cases}
\end{equation}
Finally, the base LLM is updated using this loss signal via a single optimization step (line~14).

\section{Performance Evaluation}
\label{Experiment}

To assess the efficacy of ORACL for both RCA and autoscaling in microservice environments, we conduct experiments on a realistic Kubernetes-based testbed and on open-source datasets. We detail the experimental settings, benchmarks, and provide a comprehensive analysis of the results. The primary objective is to demonstrate that these tasks can be unified within a single framework and that the CoT-based method is highly adaptable without extensive training, thereby suggesting promising directions for future research.

\subsection{Experimental Setup}
The prototype is deployed on a five-node Kubernetes cluster consisting of one master and four worker nodes, each equipped with 8 vCPUs and 8~GB of memory, with cgroup v2 enabled. Two additional GPU nodes are used for training and inference, including one node with four NVIDIA A100 GPUs and another with eight NVIDIA RTX 6000 Ada GPUs. Model training is implemented using PyTorch and DeepSpeed ZeRO-3 to support efficient distributed training through Unsloth\footnote{\url{https://github.com/unslothai/unsloth}} with quantization-based optimization. RCA experiments are conducted using the Murphy dataset~\cite{harshavardhan2023murphy} on the Hotel Reservation benchmark~\cite{gan2019deathstarbench}, while autoscaling evaluations are performed on Sock-Shop~\cite{sockshop} and Train-Ticket~\cite{gan2019deathstarbench}. Workloads are generated using a distributed Locust setup\footnote{\url{https://locust.io}} following traffic patterns derived from the Alibaba trace~\cite{luo2022power}. Performance metrics and traces are collected through Prometheus, cAdvisor, and Jaeger.

\subsection{Baselines}
Given the two tasks in this work, we evaluate our model against two groups of baselines.
\textbf{Root-cause detection.}
We compare with four representative systems: ExplainIT~\cite{jeyakumar2019explainit} (pairwise metric correlations), NetMedic~\cite{kandula2009detailed} and Sage~\cite{gan2021sage} (dependency-graph diagnosis with correlation-based edge weights and heuristics), and Murphy~\cite{harshavardhan2023murphy} (MRF-based diagnosis over loosely associated metrics). We also evaluate our constructed prompts on three commercial LLM APIs: ChatGPT-4o~\cite{openai2025gpt5}, Deepseek-V3~\cite{deepseek2025r1}, and Claude-3.5-Haiku~\cite{anthropic2025claude35haiku}.

\textbf{Resource allocation.}
For autoscaling, we consider three baselines: KuScal~\cite{kubernetes}, a threshold-based Kubernetes scaler with a 75\% CPU target as suggested in prior work~\cite{carrion2022kubernetes}; CoScale~\cite{xu2022coscal}, a correlation-driven Q-learning autoscaler with GRU-based workload prediction and support for horizontal/vertical scaling and brownout; and FIRM~\cite{qiu2020firm}, a model-based controller that allocates resources along critical microservice paths to reduce end-to-end latency.

\subsection{Root Cause Identification}

To evaluate the effectiveness of our CoT design and demonstrate the benefits of our training strategy, we first compare ORACL against several state-of-the-art RCA systems. We then apply the prompts generated by our interface to a set of commercial large-scale models, each with at least an order of magnitude more parameters, to assess the robustness and generalization of our training method.

\begin{figure}[h]
    \centering
    \includegraphics[width=0.9\linewidth]{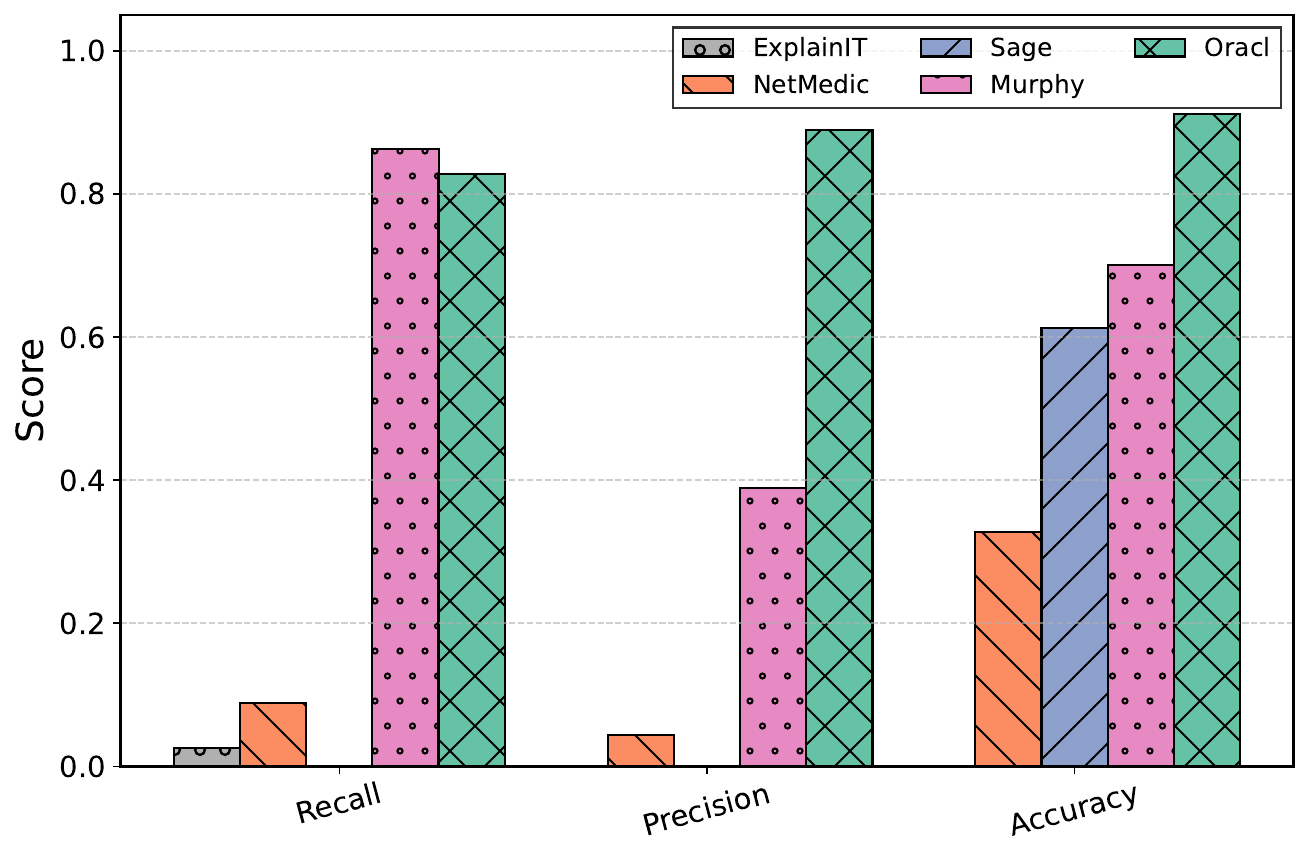}
    \caption{RCA Comparison}
    \label{fig:raccomparison}
\end{figure}

\subsubsection{Comparison with State-of-the-Art RCA Models}
In Fig.~\ref{fig:raccomparison}, we evaluate ORACL against four representative RCA frameworks, including ExplainIT, NetMedic, Sage, and Murphy on the original datasets released with Murphy, using Recall, Precision, and Accuracy. For a fair comparison, we adopt a strict protocol in which only the single most confident prediction of each method is counted.

Traditional statistical and rule-based baselines (ExplainIT, NetMedic) exhibit limited diagnostic capability and fail to capture cyclic failure conditions~\cite{harshavardhan2023murphy}, while Sage shows unstable precision and incomplete recall. Murphy, a recent learning-based method, attains slightly higher recall than ORACL but substantially lower precision and accuracy under topology churn. In contrast, ORACL achieves 0.83 recall, 0.89 precision, and 0.92 accuracy, combining high anomaly sensitivity with a low false-positive rate and demonstrating robust generalization across diverse microservice conditions.

\subsubsection{Comparison with Commercial LLMs}
Table~\ref{tab:apimetrics} summarizes the comparative performance of ORACL and three state-of-the-art commercial LLMs under the default ORACL configuration. Specifically, we compare ORACL against DeepSeek-V3, ChatGPT-4o, and Claude~3.5~Haiku. Public reports indicate that DeepSeek-V3 comprises approximately 671 billion parameters, while ChatGPT-4o is widely believed to operate in a similar trillion-scale regime. In contrast, Claude~3.5~Haiku is explicitly positioned as a lightweight model with a substantially smaller parameter budget, although exact figures are not publicly disclosed. Despite using significantly fewer parameters than all three commercial systems, ORACL consistently achieves superior performance in our setting. Quantitatively, ORACL improves precision by approximately 3\%–28\%, recall by 6\%–38\%, and accuracy by 4\%–15\% over the commercial baselines. For example, compared to ChatGPT-4o, ORACL attains gains of +6\% in precision, +6\% in recall, and +5\% in accuracy. Overall, ORACL achieves the highest precision, recall, and accuracy among all evaluated models.

\begin{table}[h]
\centering
\renewcommand{\arraystretch}{1}
\setlength{\tabcolsep}{12pt} 
\caption{Performance comparison across commercial models}
\label{tab:apimetrics}
\begin{tabular}{lccc}
\toprule
\textbf{Model} & \textbf{Precision} & \textbf{Recall} & \textbf{Accuracy} \\
\midrule
Claude-3.5-Haiku & 0.61 & 0.45 & 0.77 \\
ChatGPT-4o & 0.83 & 0.77 & 0.87 \\
DeepSeek-V3 & 0.86 & 0.72 & 0.88 \\
ORACL & \textbf{0.89} & \textbf{0.83} & \textbf{0.92} \\
\bottomrule
\end{tabular}
\end{table}

\begin{figure*}[t]
  \centering
  \subfloat[RPS Comparison\label{fig:sockshopRps}]{
    \includegraphics[width=0.32\textwidth,height=4cm,keepaspectratio]{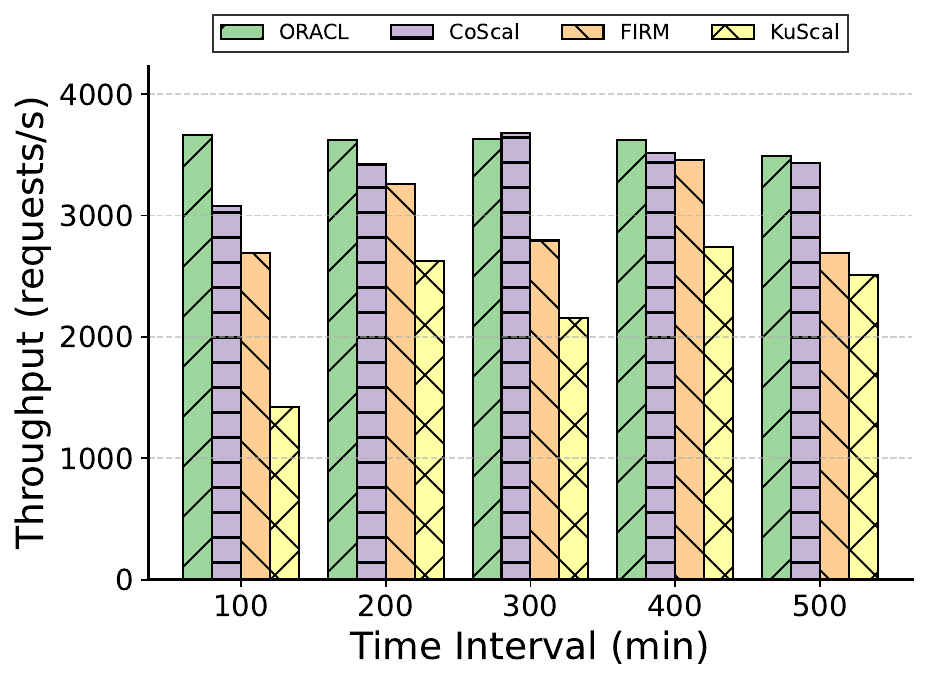}
  }\hfill
  \subfloat[Average Response Time\label{fig:sockshopAverageResponseTime}]{
    \includegraphics[width=0.32\textwidth,height=4cm,keepaspectratio]{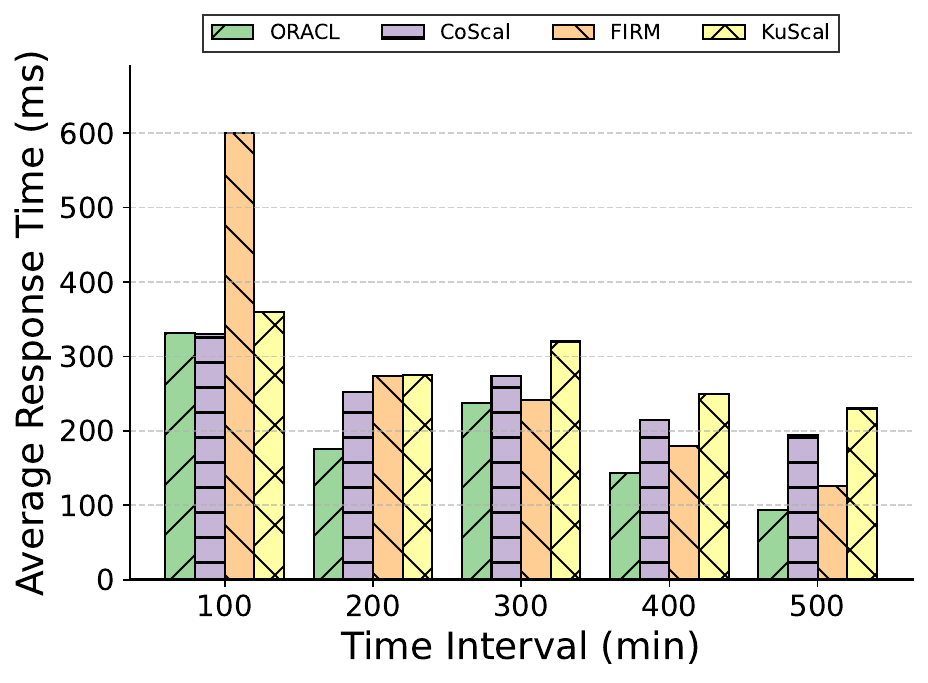}
  }\hfill
  \subfloat[Response Time CDF\label{fig:sockshopResponseTimeCDF}]{
    \includegraphics[width=0.32\textwidth,height=4cm,keepaspectratio]{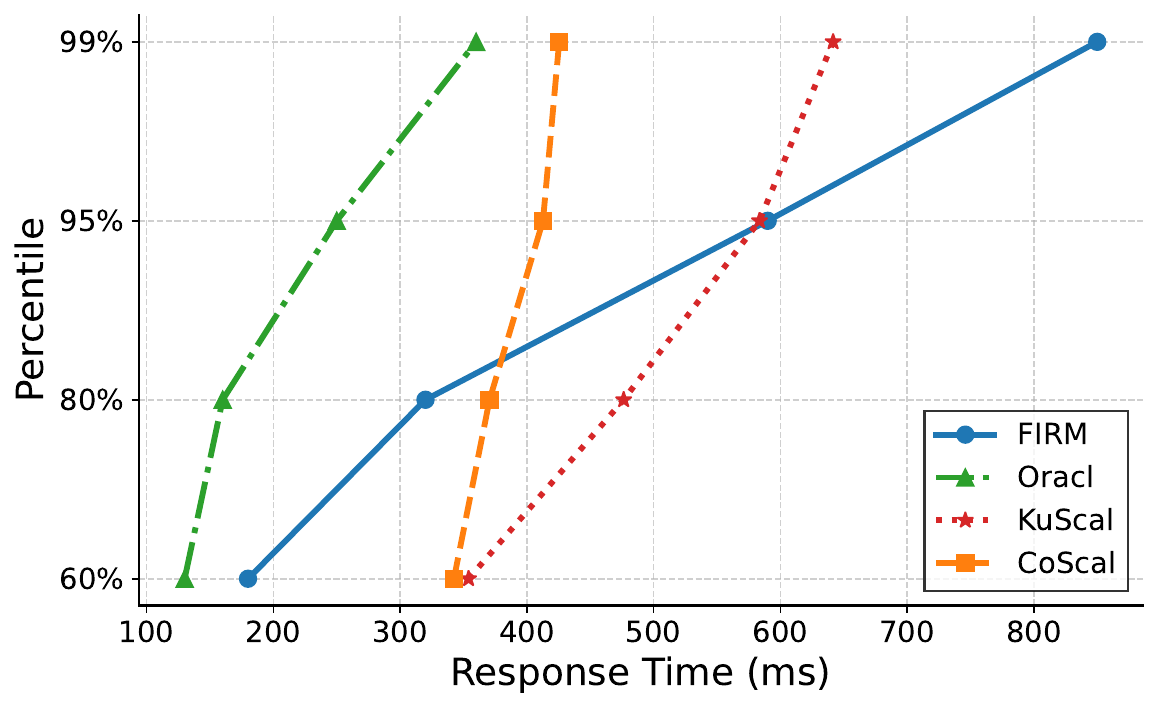}
  }
  \caption{Comparison of Sock-Shop performance metrics.}
  \label{fig:sockshop_overall}
\end{figure*}

\subsection{Scheduling Evaluations}

We evaluate ORACL under two research questions: (1) its performance relative to existing approaches under limited training, and (2) its performance when competing models are fully converged. Experiments are conducted in two scenarios, where the 500-minute and 5000-minute test traces are each divided into five equal intervals, and reported results are averaged per interval. ORACL performs a single data-collection phase of approximately 1--2 hours (Section~\ref{sec:resourceallocation}) and requires no further online retraining.

The first scenario evaluates Sock-shop with only 24 hours of training data. The second considers the larger Train-ticket system, where all models are allowed to fully converge. Due to its substantially larger call graph, Train-ticket is handled by multiple LLM-agent replicas operating on different subgraphs; in case of conflicting actions, the system conservatively selects the decision that provisions additional resources.

\subsubsection{\textbf{Requests-per-Second (RPS) Analysis}}

To evaluate the system’s throughput and the scheduler’s ability to effectively utilize available resources, we analyze the requests per second (RPS), which measures how many requests the system can successfully process within one second.

\textit{Sock-shop (limited training).}
Fig.~\ref{fig:sockshopRps} reports the RPS of the four methods on Sock-shop over five time periods. ORACL consistently achieves the highest throughput between around 3500 and 3670, delivering about 5\% more requests/s than CoScal and over 20\% more than FIRM, while KuScal lags far behind. In addition, ORACL’s throughput is also the most stable over time, whereas FIRM shows noticeable oscillations under previously unseen states and KuScal is consistently under-provisioned. With only one day of training data, the LLM-based ORACL scheduler already achieves both higher and more stable throughput than all baselines.

\textit{Train-ticket (fully converged).}
Fig.~\ref{fig:trainticketRPS} presents the RPS results on Train-ticket when all baseline methods are trained to convergence. FIRM now achieves the highest average throughput of around 750 requests/s, with ORACL and CoScal closely following at around 730 requests/s. Compared with FIRM, ORACL incurs only about 2\% lower throughput, while still improving throughput over KuScal by more than 30\%. The small gap between ORACL and the best converged deep-learning model suggests that ORACL generalizes well to a more complex application, even though it does not rely on long-term online training.

\subsubsection{\textbf{Average Response Time}}

To evaluate the system’s stability in handling real workloads with minimal failures, we further compare the average end-to-end response time across the same five periods.

\textit{Sock-shop (limited training).}
Fig.~\ref{fig:sockshopAverageResponseTime} shows that, on Sock-shop, ORACL consistently achieves the lowest latency across all five periods. Excluding the initial warm-up interval, where FIRM spikes to nearly 600ms, ORACL attains a steady-state average of 162.3ms, compared with 234.0ms for CoScal, 205.5ms for FIRM, and 268.5ms for KuScal. This corresponds to a latency reduction of approximately 31\%, 21\%, and 40\%, respectively. The pronounced first-period spike indicates that FIRM is sensitive to previously unseen workload patterns under limited training, whereas ORACL maintains low and stable latency after only a short offline data-collection phase.

\begin{figure*}[t]
  \centering
  \subfloat[RPS Comparison\label{fig:trainticketRPS}]{
    \includegraphics[width=0.32\textwidth,height=4cm,keepaspectratio]
      {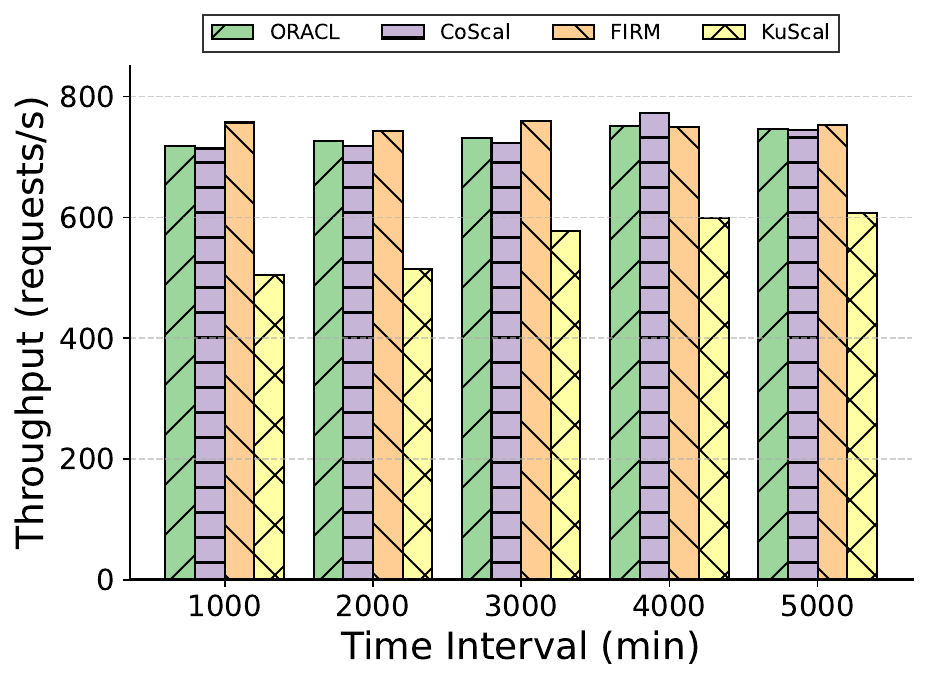}
  }\hfill
  \subfloat[Average Response Time\label{fig:trainticketAverageResponseTime}]{
    \includegraphics[width=0.32\textwidth,height=4cm,keepaspectratio]
      {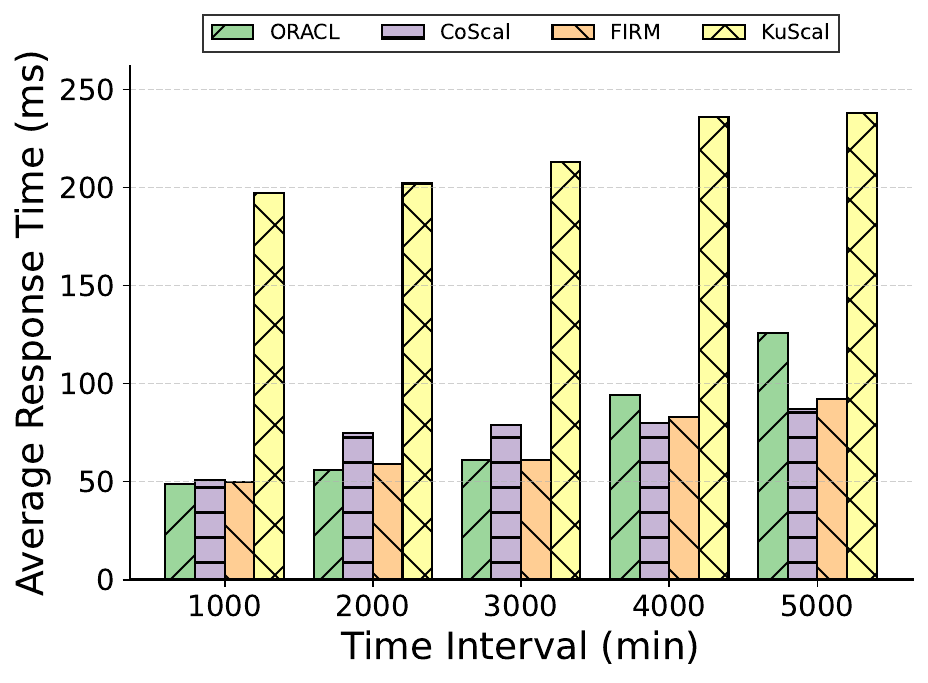}
  }\hfill
  \subfloat[Response Time CDF\label{fig:trainticketCDF}]{
    \includegraphics[width=0.32\textwidth,height=4cm,keepaspectratio]
      {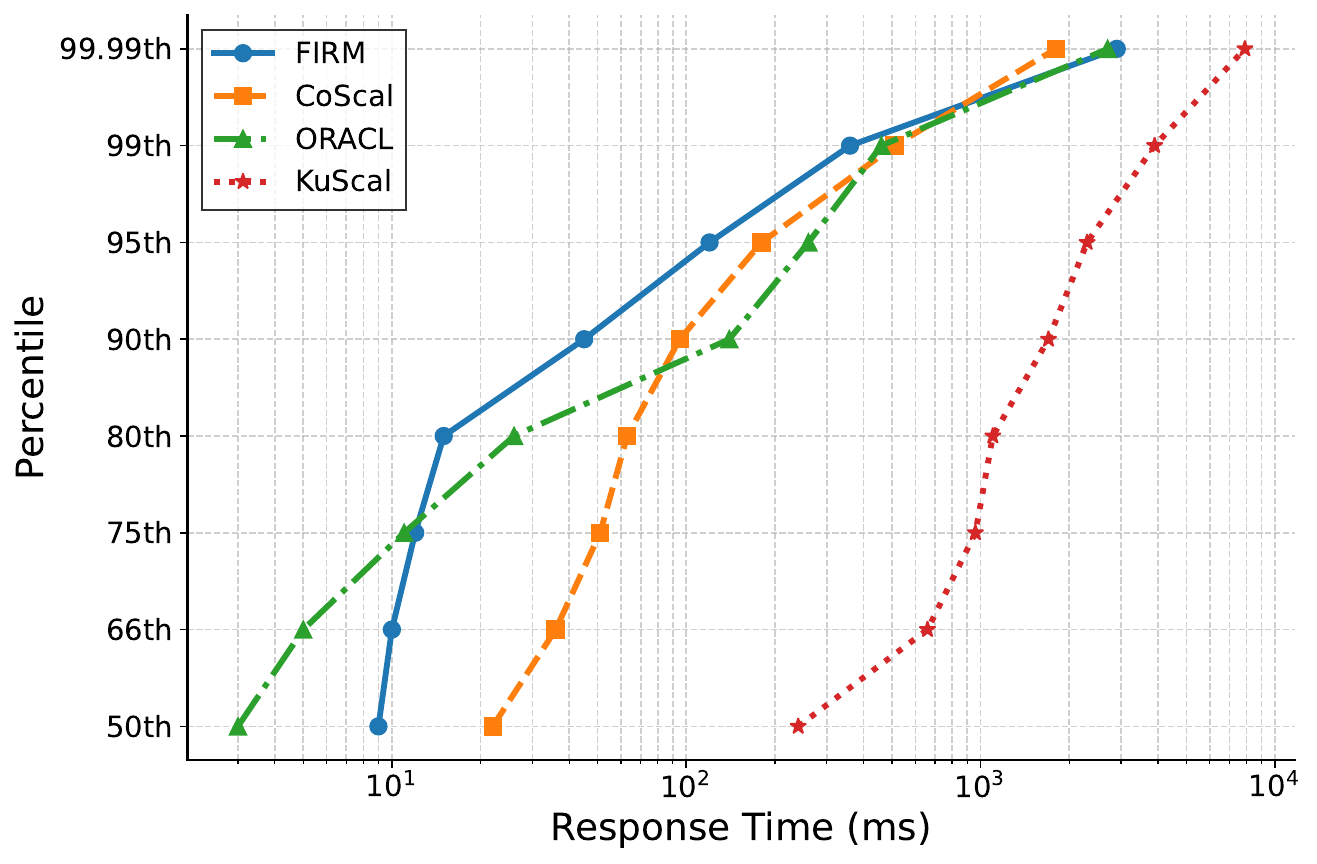}
  }
  \caption{Comparison of Train-Ticket performance metrics.}
  \label{fig:trainticket_overall}
\end{figure*}

\textit{Train-ticket (fully converged).
}The response-time results for Train-ticket are summarized in Fig.~\ref{fig:trainticketAverageResponseTime}. All three advanced methods (FIRM, CoScal, and ORACL) achieve low latencies on the order of tens of milliseconds. Compared with the best deep-learning baseline (FIRM), ORACL incurs about 10\% higher latency and only a few percent overhead compared with CoScal, but still reduces latency by roughly 65\% compared with KuScal. Looking at the per-period behavior, ORACL achieves the lowest or comparable latency in the first three periods, where the workload is relatively stable. In the last two periods, its latency increases more noticeably, while FIRM and CoScal remain slightly lower. This suggests that, in the presence of long-running shifts in user behavior or carrier composition, the converged numerical models can capture fine-grained patterns more accurately, whereas ORACL trades some optimality for its lightweight training.

\subsubsection{\textbf{Cumulative Distribution Function (CDF) of Response Time}}
To evaluate the effectiveness of resource usage and better characterize system behavior, we further analyze the response-time CDFs to reveal how each scheduler behaves across the full latency distribution.

\textit{Sock-shop (limited training).}
Fig.~\ref{fig:sockshopResponseTimeCDF} shows the response-time CDF on Sock-shop. ORACL achieves the tightest distribution, reducing latency by about 30--50\% at the 60th--80th percentiles and by 15--60\% at the tail (95th--99th percentiles) relative to the baselines. Among the non-ORACL baselines, FIRM performs best at lower quantiles but exhibits the heaviest tail, indicating instability under unseen workloads, while CoScal and KuScal display smoother yet consistently higher latencies. These results indicate that ORACL effectively stabilizes tail behavior under limited training by prioritizing QoS-critical services.

\textit{Train-ticket (fully converged).}
Fig.~\ref{fig:trainticketCDF} presents the response-time CDFs on Train-ticket. ORACL clearly dominates at low percentiles, with median and 66th-percentile latencies reduced by roughly 50--67\% relative to FIRM and over 80\% compared with CoScal, indicating its strong effectiveness in optimizing the critical path of short requests. ORACL remains competitive around the 75th percentile and, although FIRM and CoScal slightly outperform it in the mid-to-high range (80th--95th percentiles) after full convergence, ORACL matches FIRM near the tail and significantly outperforms KuScal, reducing worst-case latency by about two-thirds. This behavior stems from ORACL’s design: the LLM-based scheduler prioritizes reasoning about the root causes of QoS violations rather than aggressively reclaiming resources from all services, thereby preserving capacity on non-critical microservices. While this conservative strategy can yield marginally higher mid-percentile latency, it consistently delivers strong median performance and substantially improved tail behavior.

\subsubsection{\textbf{Summary}}

Overall, the threshold-based CoScal provides reasonable performance and convergence speed, but its coarse state discretization and limited action space prevent it from fully exploiting the available resources, making it the most ``unremarkable'' method across all ten periods. The LLM-based, chain-of-thought ORACL scheduler is the most adaptive when the application or workload changes: in both Sock-shop and Train-ticket, it maintains stable behavior without per-application hyperparameter tuning and achieves strong throughput and latency with only a short data-collection phase.

Deep-learning-based FIRM, on the other hand, benefits from large amounts of training data. Under limited training (Sock-shop), it suffers from instability and occasional severe latency spikes, but once fully converged on a complex workload (Train-ticket), it can slightly surpass ORACL in both throughput and latency. In summary, ORACL offers a favorable trade-off: it avoids the long training time and instability of deep-learning approaches while achieving near state-of-the-art performance on a complex microservice system and significantly outperforming Kubernetes-based baselines.
\begin{table}[!t]
\centering
\caption{CDF Comparison in Train-ticket}
\label{tab:percentiles}
\setlength{\tabcolsep}{3.5pt} 
\begin{tabular}{lcccccccc}
\toprule
\textbf{System} & 50th & 66th & 75th & 80th & 90th & 95th & 99th & 99.99th \\
\midrule
FIRM   & 9  & 10 & 12 & \textbf{15}  & \textbf{45}  & \textbf{120} & \textbf{360} & 2900 \\
CoScal & 22 & 36 & 51 & 63  & 95  & 180 & 510 & \textbf{1800} \\
KuScal & 240 & 660 & 960 & 1100 & 1700 & 2300 & 3900 & 7900 \\
ORACL & \textbf{3}  & \textbf{5}  & \textbf{11} & 26  & 140 & 260 & 460 & 2700 \\
\bottomrule
\end{tabular}
\end{table}


\subsection{Attention Mechanism Analysis}
\begin{figure}[!t]
  \centering

  \begin{minipage}[t]{0.32\linewidth}
    \centering
    \subfloat[Layer 64, Head 35\label{fig:l64h35Attention}]{
      \includegraphics[page=1,width=\linewidth]{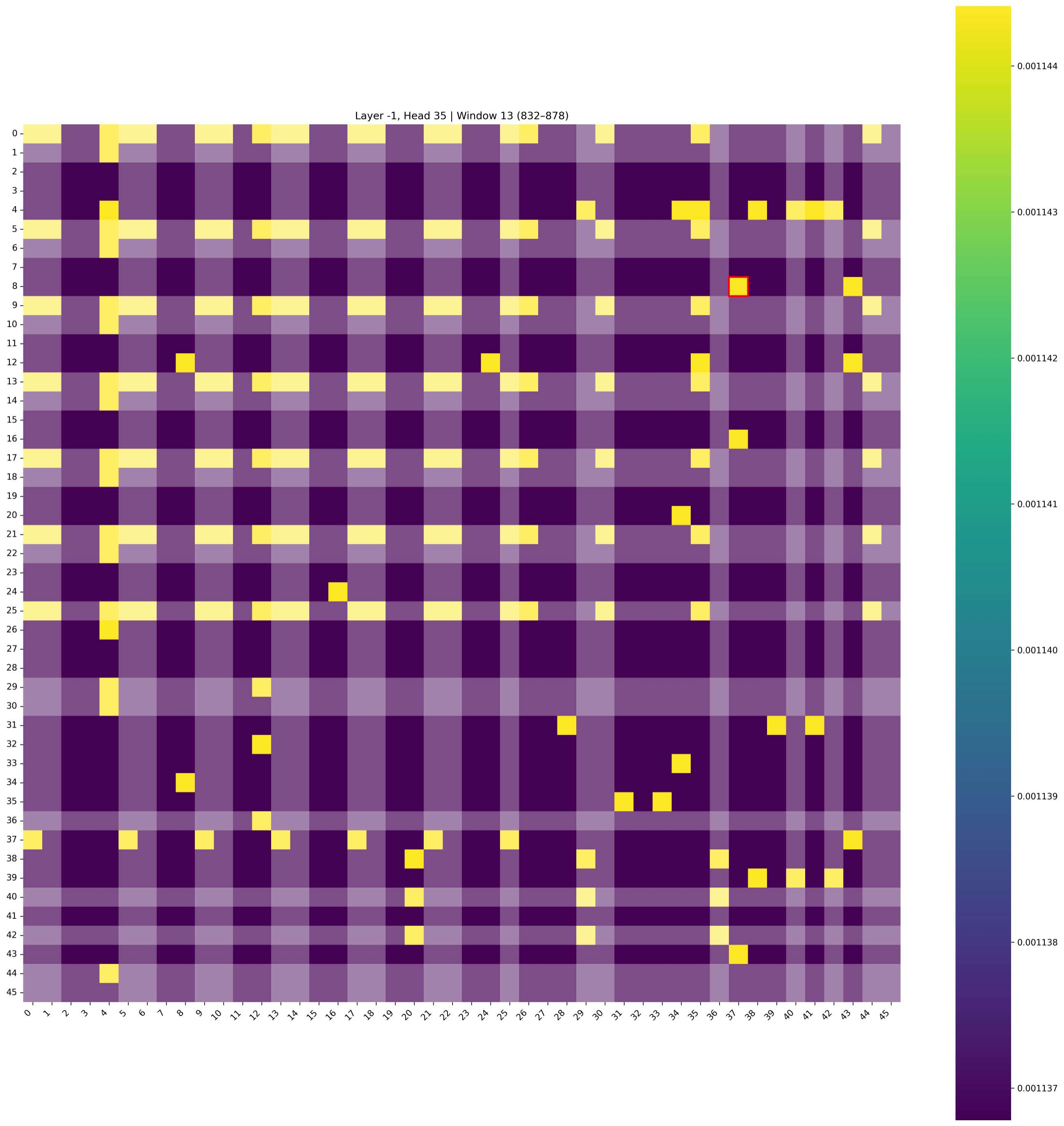}
    }
  \end{minipage}\hfill
  \begin{minipage}[t]{0.32\linewidth}
    \centering
    \subfloat[Layer 2, Head 21\label{fig:l2h21Attention}]{
      \includegraphics[page=2,width=\linewidth]{pics/Attention.pdf}
    }
  \end{minipage}\hfill
  \begin{minipage}[t]{0.32\linewidth}
    \centering
    \subfloat[Layer 25, Head 2\label{fig:l25h2Attention}]{
      \includegraphics[page=3,width=\linewidth]{pics/Attention.pdf}
    }
  \end{minipage}

\par\medskip

  \begin{minipage}[t]{0.48\linewidth}
    \centering
    \subfloat[Layer 29, Head 15\label{fig:l29h15Attention}]{
      \includegraphics[page=4,width=\linewidth]{pics/Attention.pdf}
    }
  \end{minipage}\hfill
  \begin{minipage}[t]{0.48\linewidth}
    \centering
    \subfloat[Layer 29, Head 10\label{fig:l29h10Attention}]{
      \includegraphics[page=5,width=\linewidth]{pics/Attention.pdf}
    }
  \end{minipage}

  \caption{Attention Head Visualization Across Different Layers.}
  \label{fig:attention_pages}
\end{figure}

To better understand the model’s internal reasoning and explainability, we analyze token-wise attention patterns across early, middle, and late layers of the trained model. Tokens that most strongly influence the final prediction are highlighted in red, with color intensity indicating attention magnitude. Although it is impractical to visualize all tokens involved in the reasoning process, inspection of representative attention patterns reveals that jointly modeling optimization and root-cause identification through the CoT process is both effective and interpretable. Despite the large token scale inherent to LLM inference, the critical patterns remain clearly observable and align with prior findings in NLP research.

In the early layers, as depicted in Fig.~\ref{fig:l2h21Attention}, the model primarily focuses on capturing local contextual and syntactic relations. At this stage, the attention patterns mainly highlight tokens that represent resource-related words, reflecting how the model learns word-level and adjacent dependencies, similar to conventional NLP models. Since the special tokens introduced in our setup carry limited semantic meaning, it is reasonable to observe lower attention scores on the left side of the figure.

The middle layers capture long-term dependencies and causal patterns between microservices, assigning closer attention to token pairs representing pod-level interactions on the same call graph. Metric–behavior alignment is also evident, as shown by the red-highlighted regions corresponding to Front-end–Memory (Fig.~\ref{fig:l29h10Attention}), Rate–CPU (Fig.~\ref{fig:l29h15Attention}), and Rate–Memory (Fig.~\ref{fig:l25h2Attention}).

In the later stage, particularly in the final layer (as shown in Fig.~\ref{fig:l64h35Attention}), the model focuses more on the structural patterns of the predefined templates to generate plans interpretable by our framework. Moreover, as the LLM is optimized for next-token prediction, some token pairs exhibit elevated attention beyond structural alignment. While such peaks may appear to reflect causal reasoning, they more likely indicate output alignment toward correct root-cause predictions rather than explicit causal analysis.

\section{Conclusions and Future Work}\label{sec:Limitation}

Our proposed ORACL system unifies root-cause detection and resource allocation tasks within a single framework. Compared with existing techniques, it requires significantly shorter training time while still achieving competitive performance in practical settings. Furthermore, most of the workflow is specified at a high level, which reduces the overall coding effort. While our results demonstrate the feasibility of optimizing chain-of-thought reasoning for resource allocation and root causes finding, there exist several opportunities for further improvement as discussed below:

\begin{itemize}
   \item \textbf{Inference efficiency.} Although inference is generally fast on average, the high computational and energy cost of large language models, together with non-negligible and variable inference latency, makes end-to-end responsiveness sensitive to request frequency and workload bursts. In several experiments, inference occasionally exceeded the allocated time window, causing transient stalls. Future work will explore more efficient, domain-adapted models and optimized serving infrastructures to reduce inference cost, stabilize latency, and improve robustness under peak load.

    \item \textbf{Training data coverage.} The current dataset spans a limited range of microservice architectures and deployment environments, which restricts generalization. More diverse and systematically collected datasets are needed to capture a broader spectrum of failure patterns, workload characteristics, cloud platforms, and operational factors such as network conditions and workload dynamics. In addition, the lack of energy-efficiency–oriented datasets limits the model’s ability to reason about sustainability-aware resource allocation.

    \item \textbf{Granularity of analysis.} Reasoning is primarily performed at the service and resource levels. Future work could leverage retrieval-augmented generation (RAG) to construct structured, time-aware telemetry representations that capture system-state relationships, by integrating node-level scheduling information, container-level contention signals, and code-level efficiency metrics, thereby enabling finer-grained diagnoses and more precise allocation decisions.
    
\end{itemize}

Despite these limitations, our experiments suggest that direct CoT optimization is a promising direction. With richer data, tighter integration with cloud-native tooling, and improved model architectures, LLM-based agents could evolve into autonomous controllers for cloud resource management.

\bibliographystyle{IEEEtran}
\bibliography{references}

\vspace{-3.5\baselineskip}
\begin{IEEEbiography}[{\includegraphics[width=1in,height=1.25in,clip,keepaspectratio]{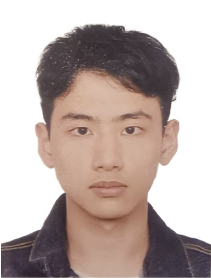}}]{Haoyu Bai}
received his B.Sc. and Master's degree from the University of Melbourne, Australia. He is currently pursuing Ph.D. degree with the School of Computing and Information Systems, University of Melbourne. His research interests include efficient microservice-based applications, resource management, and cloud system optimization.
\end{IEEEbiography}

\vspace{-2.0\baselineskip}
\begin{IEEEbiography}[{\includegraphics[width=1in,height=1.25in,clip,keepaspectratio]{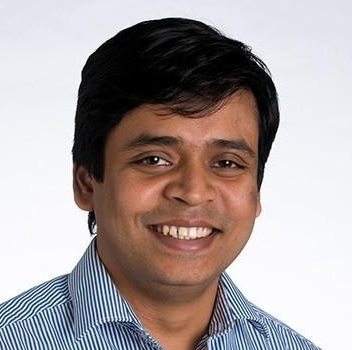}}]{Muhammed Tawfiqul Islam}
is a Lecturer in the School of Computing and Information Systems at the University of Melbourne. He received his Ph.D. in Engineering from the University of Melbourne in 2021, working in the CLOUDS Laboratory. His doctoral research on scalable cloud resource management received the IEEE TCSC Outstanding Ph.D. Dissertation Award in 2021. His research interests span distributed systems, cloud and edge computing, microservice scheduling,  and performance and cost-efficient resource management.
\end{IEEEbiography}

\vspace{-1\baselineskip}
\begin{IEEEbiography}[{\includegraphics[width=1in,height=1.25in,clip,keepaspectratio]{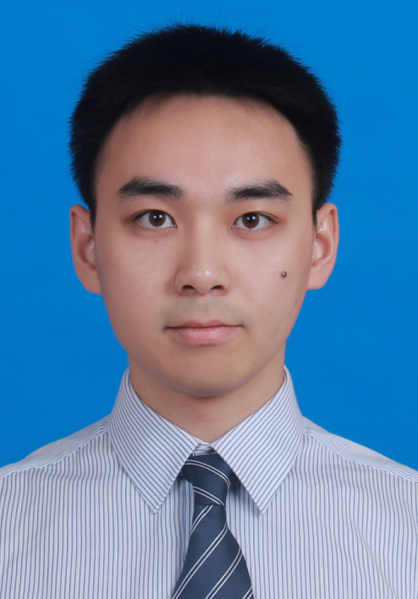}}]{Minxian Xu}
is an Associate Professor with the Shenzhen Institute of Advanced Technology, Chinese Academy of Sciences. He received his Ph.D. degree from the University of Melbourne in 2019. His research interests include resource scheduling, performance optimization, infrastructure for large language models, and cloud computing systems. He has co-authored more than 80 peer-reviewed papers in leading international conferences and journals. His Ph.D. thesis received the 2019 IEEE TCSC Outstanding Ph.D. Dissertation Award, and he was the recipient of the 2023 IEEE TCSC Award for Excellence (Early Career Award).
\end{IEEEbiography}

\vspace{-1\baselineskip}
\begin{IEEEbiography}[{\includegraphics[width=1in,height=1.25in,clip,keepaspectratio]{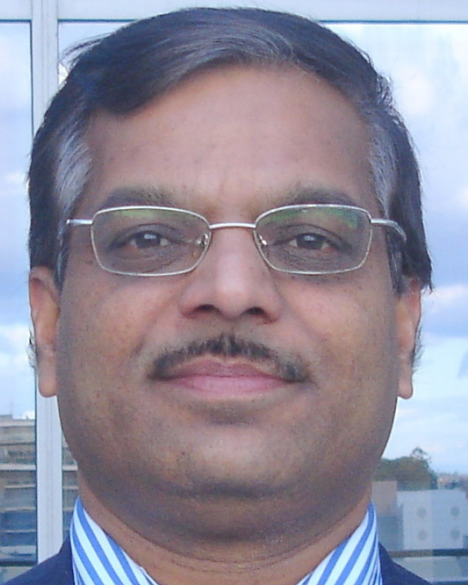}}]{Rajkumar Buyya}
is a Redmond Barry Distinguished Professor and Director of the Quantum Cloud Computing and Distributed Systems (qCLOUDS) Laboratory at the University of Melbourne, Australia. He has authored more than 850 publications and seven textbooks, including Mastering Cloud Computing published by McGraw-Hill, China Machine Press, and Morgan Kaufmann. Recognized as one of the world’s most highly cited researchers in computer science and software engineering (h-index: 178; g-index: 394; 168,800+ citations), Dr. Buyya is a Fellow of IEEE, a Foreign Fellow of Academia Europaea, and a Fellow of ACM. He co-founded five major IEEE/ACM international conferences—CCGrid, Cluster, Grid, e-Science, and UCC—and served as the Chair of their inaugural meetings. He served as the founding Editor-in-Chief of the IEEE Transactions on Cloud Computing. He is currently serving as Co-Editor-in-Chief of Journal of Software: Practice and Experience, which was established 55+ years ago.

\end{IEEEbiography}

\end{document}